\documentclass{ws-ijam}
\usepackage[square]{natbib}
\usepackage[colorlinks=true,allcolors=black]{hyperref}

\begin{document}

\markboth{D.~Y.~Wang et al}{Multiple Steady Oscillatory Solutions in a Collapsible Channel Flow}

\catchline{}{}{}{}{}

\title{Multiple Steady and Oscillatory Solutions in a Collapsible Channel Flow}

\author{Danyang~Wang\footnote{danyang.wang@glasgow.ac.uk}, Xiaoyu~Luo, and Peter~S.~Stewart}

\address{School of Mathematics and Statistics, University of Glasgow, Glasgow, G12 8SQ UK}

\maketitle

\begin{history}
\received{date}
\accepted{date}
\end{history}

\begin{abstract}
We study flow driven through a finite-length planar rigid channel by a fixed upstream flux, where a segment of one wall is replaced by a pre-stressed elastic beam subject to uniform external pressure. The steady and unsteady systems are solved using a finite element method. Previous studies have shown that the system can exhibit three steady states for some parameters (termed the upper, intermediate and lower steady branches, respectively). Of these, the intermediate branch is always unstable while the upper and lower steady branches can (independently) become unstable to self-excited oscillations. We show that for some parameter combinations the system is unstable to both upper and lower branch oscillations simultaneously. However, we show that these two instabilities eventually merge together for large enough Reynolds numbers, exhibiting a nonlinear limit cycle which retains characteristics of both the upper and lower branches of oscillations.
Furthermore, we show that increasing the beam pre-tension suppresses the region of multiple steady states but preserves the onset of oscillations. Conversely, increasing the beam thickness (a proxy for increasing bending stiffness) suppresses both multiple steady states and the onset of oscillations.
\keywords{flow-vessel interaction; collapsible channel flows; stability.}
\end{abstract}

\section{Introduction}
\label{sec:intro}	

Fluid flow through flexible conduits in the human body can exhibit a wide variety of interesting physiological phenomena \citep{Heil&Hazel2011}; such flows can be investigated experimentally using a Starling Resistor, where fluid is driven through a segment of externally pressurised flexible tubing by either a prescribed driving pressure or a prescribed upstream flux \citep[e.g.~][]{bertram1982two,bertram1986unstable}. In particular, pressure driven flow can exhibit a phenomenon known as `flow limitation', where the flow rate along the tube does not continue to increase as the driving increases: the large flow speeds collapse the tube through the Bernoulli effect inhibiting the flow \citep{bertram1999flow,bertram2006onset}. Similarly, flux driven systems can exhibit the associated phenomenon of `pressure drop limitation', where the pressure difference between the ends of the collapsible tube does not increase with increases in flow rate \citep{bertram1986unstable,bertram1990mapping}. In physiology, flow limitation can occur during tidal breathing, when the flow rate of air being expelled from the lung becomes maximised during forced expiration \citep{tantucci2013expiratory}. 

In some cases these highly collapsed steady configurations can co-exist with other, more inflated, configurations of the vessel. Such multiplicity in steady solutions is evident in Starling Resistor experiments, where the system can exhibit hysteresis i.e.~different steady configurations at the same operating point dependent on the history and termed an `open-to-closed transition' \citep{bertram1991application,bertram1999flow}. Such co-existing steady states are also evident in theoretical models of flow in collapsible tubes \citep{hazel2003steady,heil2010self}. In particular, lumped parameter models indicate that the collapsible tube can typically exhibit three co-existing branches of steady solutions across a range of parameters \citep{armitstead1996study}; these three branches have been termed the upper branch (where the tube is mostly inflated), the lower branch (where the tube is highly collapsed) with an intermediate branch between them. The upper (lower) steady solutions is connected to the intermediate branch through a saddle-node bifurcation at the upper (lower) branch limit point; the upper and lower steady branches are stable to non-oscillatory perturbations and the intermediate branch is always unstable  \citep{armitstead1996study}.

In physiology, flow limitation in the lung airways is often accompanied by wheezing, associated with rapid fluttering of the airway wall \citep{grotberg1989flutter}. Similarly, Starling Resistor experiments investigating flow limitation can sometimes exhibit spontaneous self-excited oscillations which fall into distinct frequency bands \citep{bertram1982mathematical,bertram1990mapping,bertram1991application}. Such self-excited oscillations are also evident in theoretical models of flow in collapsible tubes, such as lumped parameter models \citep{bertram1982mathematical,armitstead1996study}, cross-sectionally averaged one-dimensional models \citep{jensen1990instabilities} and full three-dimensional models \citep{heil2010self,whittaker2010predicting}. 

In this study we consider a theoretical model for a planar analog of the Starling Resistor experiment, formed by removing a segment of one wall of a rigid channel and inserting an elastic wall. This planar analog exhibits multiple steady solutions in some parameter regimes \citep{luo2000multiple,heil2004efficient,stewart2017instabilities,wang2021energetics,herrada2021global}, with a three branch structure qualitatively similar to the collapsible tube models. In addition, this collapsible channel system exhibits transition to self-excited oscillations from both the lower (collapsed) branch of steady solutions \citep{heil2004efficient,stewart2017instabilities} as well as the upper (inflated) branch of steady solutions \citep{herrada2021global,wang2021energetics}. 
Fully developed upper branch oscillations exhibit an upstream propagating hump along the compliant segment; this wave is suppressed by reflection at the upper rigid segment and replaced by a new wave at the downstream end of the compliant segment \citep{wang2021energetics}. By contrast, fully developed lower branch oscillations exhibit a constriction at the downstream end of the compliant segment which propagates up and downstream over a narrow range \citep{luo1996numerical,wang2021energetics}.

In particular, we use a model where the wall takes the form of a thin pre-stressed (massless) elastic beam with resistance to both bending and stretching (Sec.~\ref{sec:model}) \citep{cai2003fluid,luo2008cascade,wang2021energetics}. We use this model to explore how the size of the region of multiple steady states is influenced by the system parameters, particularly the beam pre-tension and bending stiffness (Sec.~\ref{sec:Mult_st}). Finally, we also use this model to explore the onset of self-excited oscillations from both the upper and lower branches of steady solutions, showing that in some cases these two families of oscillations merge together (Sec.~\ref{sec:unsteady}).

\section{The Model}\label{sec:model}
\begin{figure}[t!]
\centering
\includegraphics[width=0.8\textwidth]{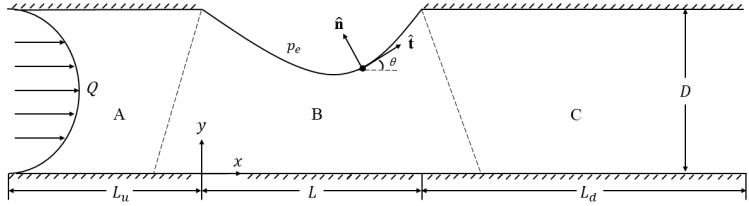}
\caption{The fluid-beam model.}
\label{Fig:BemMod}
\end{figure}
We consider an incompressible Newtonian fluid (of density $\rho$ and viscosity $\mu$) driven through a finite-length rigid channel with uniform width $D$. 
We consider a parabolic inlet flow driven through the channel with flux $Q$ against a fixed outlet pressure $p_0$. 
One segment of the upper wall is replaced by a plane strained elastic beam of thickness $h$ subject to a uniform external pressure $p_e$. The corresponding lengths of the upstream and downstream rigid and elastic segments are denoted $L_u$, $L_d$ and $L$, respectively. We parameterize the domain using a Cartesian coordinate system with origin at the intersection between the upstream rigid segment and the compliant segment on the entirely rigid wall (see Fig. \ref{Fig:BemMod}). Time is denoted $t$.

We model the flexible wall as a massless elastic beam and denote the axial pre-tension along the beam as $T$. The extensional and bending stiffnesses of the beam are denoted as $EA$ and $EJ$, respectively. Here $E$ is the Young's modulus of the material while $A$ and $J$ are the cross-sectional area and the second moment of inertia of the cross-sectional area of the beam, respectively. The undeformed elastic beam is parameterized by the coordinate $l$ (where $0\leqslant l\leqslant L$).

\subsection{Governing Equations}\label{sec: goveq}

To form non-dimensional variables we scale all lengths on $D$, velocities on $Q/D$, time on $Q/D^2$ and pressures on the inertial scale according to 
\begin{align}
p=\frac{\rho Q^2}{D^2}\tilde p+p_0,
\end{align}
where $p$ ($\tilde p$) denotes the dimensional (dimensionless) fluid pressure.

The corresponding dimensionless parameters take the form
\begin{align}
\tilde c_\lambda&=\frac{(EA)D}{\rho Q^2 }, & \tilde c_\kappa&=\frac{EJ}{\rho Q^2 D}, & \tilde T&=\frac{TD}{\rho Q^2}, & \tilde Re&=\frac{Q\rho}{\mu}, & \tilde{p}_e&=\frac{(p_e-p_0) D^2}{\rho Q^2},
\end{align}
where $\tilde c_\lambda$,  $\tilde c_\kappa$ and $\tilde T$ are the dimensionless extensional, bending stiffnesses and pre-tension of the elastic beam, respectively, $\tilde Re$ is the Reynolds number and $\tilde{p}_e$ is the external pressure.
The corresponding dimensionless lengths of the upstream, downstream and collapsible segments of the channel and the dimensionless beam thickness are scaled as
\begin{align}
\tilde L_u&=\frac{L_u}{D},&\tilde L_d&=\frac{L_d}{D},&\tilde L&=\frac{L}{D},&\tilde h&=\frac{h}{D},
\end{align}
respectively. We henceforth focus on the dimensionless quantities and drop the tildes for simplicity.

\subsubsection{Fluid Equations}
The governing equations for the (two-dimensional) fluid are the incompressible Navier-Stokes equations in the form,
\begin{equation}
\nabla\cdot{\mathbf u}=0,\quad\frac{\partial{\mathbf u}}{\partial t}+\left({\mathbf u}\cdot\nabla\right){\mathbf u}=\nabla\cdot{\boldsymbol{\sigma}},\label{FGvEq}
\end{equation}
where ${\mathbf u}=(u,v)$ is the planar fluid velocity and the Newtonian stress tensor ${\boldsymbol{\sigma}}$ takes the form
\begin{equation}
{\boldsymbol{\sigma}}=-p{\mathbf I}+Re^{-1}\left(\nabla{\mathbf u}+{\nabla{\mathbf u}}^{\rm{T}}\right),\label{eq:sigma}
\end{equation}
where  $\mathbf I$ is the identity matrix and the superscript $^{\rm{T}}$ represents the matrix transpose.

\subsubsection{Beam Equations}
We denote the two components of beam deformation as ${\mathbf x_b}=(x_b(l,~t),y_b(l,~t))$ in terms of the beam coordinate $l$; using a modified constitutive law for the massless beam, the dimensionless governing equations for the beam can be written as (details see \cite{wang2021energetics})
\begin{eqnarray}
c_{\kappa}\kappa\frac{\partial \left(\lambda\kappa\right)}{\partial l}+c_{\lambda}\frac{\partial \lambda}{\partial l}+\lambda{\sigma_1}=0,\label{lBGvEq:1}\\
-c_{\kappa}\frac{\partial }{\partial l}\left(\frac{1}{\lambda}\frac{\partial(\lambda\kappa)}{\partial l}\right)+c_{\lambda}\lambda\kappa(\lambda-1)+\lambda\kappa T+\lambda{\sigma_2}-\lambda p_e=0,\label{lBGvEq:2}\\
\frac{\partial x_b}{\partial l}=\lambda\cos\theta,\quad\frac{\partial y_b}{\partial l}=\lambda\sin\theta,\quad\frac{\partial \theta}{\partial l}=\lambda\kappa,\label{lBGvEq:3}
\end{eqnarray}
where  $\theta$ is the angle between the rigid wall and the tangent vector of the deformed beam (see Fig. \ref{Fig:BemMod}), $\kappa$ is the curvature of the beam and $\lambda$ is the principal stretch of the beam, which can each be computed in terms of the beam deformation as
\begin{align}
\kappa&=\frac{1}{\lambda^3}\left(\frac{\partial x_b}{\partial l}\frac{\partial^2y_b}{\partial l^2}-\frac{\partial y_b}{\partial l}\frac{\partial^2x_b}{\partial l^2}\right),&\lambda&=\sqrt{\left(\frac{\partial x_b}{\partial l}\right)^2+\left(\frac{\partial y_b}{\partial l}\right)^2},&(0\leq l\leq L).
\end{align}
In addition, ${\sigma_1}$ and  ${\sigma_2}$ are the tangent and normal components of the fluid traction on the beam,
\begin{equation}
\sigma_1=\left(-{\boldsymbol\sigma}\hat{{\mathbf n}}\right)\cdot\hat{{\mathbf t}},\quad\sigma_2=\left(-{\boldsymbol\sigma}\hat{{\mathbf n}}\right)\cdot\hat{{\mathbf n}}, 
\end{equation}
where $\hat{{\mathbf t}}$ and $\hat{{\mathbf n}}$ are the tangent and normal unit vectors of the deformed beam (see Fig. \ref{Fig:BemMod}).

\subsubsection{Boundary Conditions}
We prescribe a parabolic inlet flow with unit flux in the form 
\begin{align}
u&=6y(1-y), &v&=0, &(x&=-L_u, 0\leqslant y \leqslant 1).
\end{align} 
We assume the no-slip condition along the rigid walls as well as continuity of velocity between the elastic beam and the fluid in the form
\begin{eqnarray}
{\mathbf u}&={\mathbf 0},~~\quad&(y=0;~~y=1, -L_u\leq x\leq 0,~ L\leq x\leq L+L_d),\label{FluBoCon:1}\\
{\mathbf u}&={\mathbf u_{b}}=\frac{\partial {\mathbf x_b}}{\partial t},\quad&({\mathbf x}\in\partial\Omega_b).\label{FluBoCon:2}
\end{eqnarray}
The two ends of the elastic beam are fixed to the rigid wall in the form
\begin{align}
x_b(0,t)=0,\quad y_b(0,t)=1,
\quad x_b(L,t)=L,\quad y_b(L,t)=1,\quad \theta(0,t)=\theta(L,t)=0.\label{BBodCon:1}
\end{align}

\subsection{The Finite Element Method}
\label{sec:B_NumMethod}
A finite-element method is used to solve the coupled fluid-beam system.
We divide the fluid domain into three sections, denoted as A, B and C for the upstream, compliant and downstream compartments (Fig. \ref{Fig:BemMod}), respectively \citep{luo1996numerical,luo2008cascade}. 
In section B, we use an adaptive mesh that consists of  rotational spines that connect fixed nodes in the bottom wall with nodes in the deformable beam \citep{cai2003fluid}. Then nodes are seeded along these spines covering region B. Each spine can rotate around its fixed node on the rigid wall. Hence, nodes in section B can move along the rotational spine as the beam is deformed.
Further details of the numerical method are provide elsewhere \citep{luo2008cascade,hao2016arnoldi}. A mesh of 36657 elements is used for the numerical solutions in this study with time-step $\Delta t=0.01$. Convergence tests of grid- and time-step-independence were carried out between three meshes and two choices of time-step (for details see \cite{wang2021energetics}).

\subsection{Parameter Choices}\label{sec:parameter}
In our previous studies we modeled the elasticity of the beam by varying the extensional stiffness $c_{\lambda}$ \citep{luo2008cascade,wang2021energetics}, which is proportional to the bending stiffness of the beam since $c_{\kappa}=\left(h^2/12\right)c_{\lambda}$, and fixed the other beam parameters including the pre-tension ($T=0$), beam thickness ($h=0.01$) and external pressure ($p_e=1.95$). Conversely, in this study we fix $c_{\lambda}=1600$ and examine the role of these other parameters on both the steady (Sec. \ref{sec:Mult_st}) and unsteady (Sec. \ref{sec:unsteady}) behaviour of the system.
In particular we note that increasing the beam thickness allows us to alter the bending stiffness of the structure independently of the extensional stiffness.
Following \citet{luo2008cascade}, we fix the geometry of the channel according to $L_u=L=5$, $L_d=30$.

\section{Multiple Steady Solutions}\label{sec:Mult_st}
We discuss the predictions of the steady system considering the role of increasing pre-tension (Sec. \ref{sec:st-T}) and beam thickness (Sec. \ref{sec:st-ck}).

\subsection{Role of Pre-tension}\label{sec:st-T}

\begin{figure}[t!]
\centering
\includegraphics[width=\textwidth]{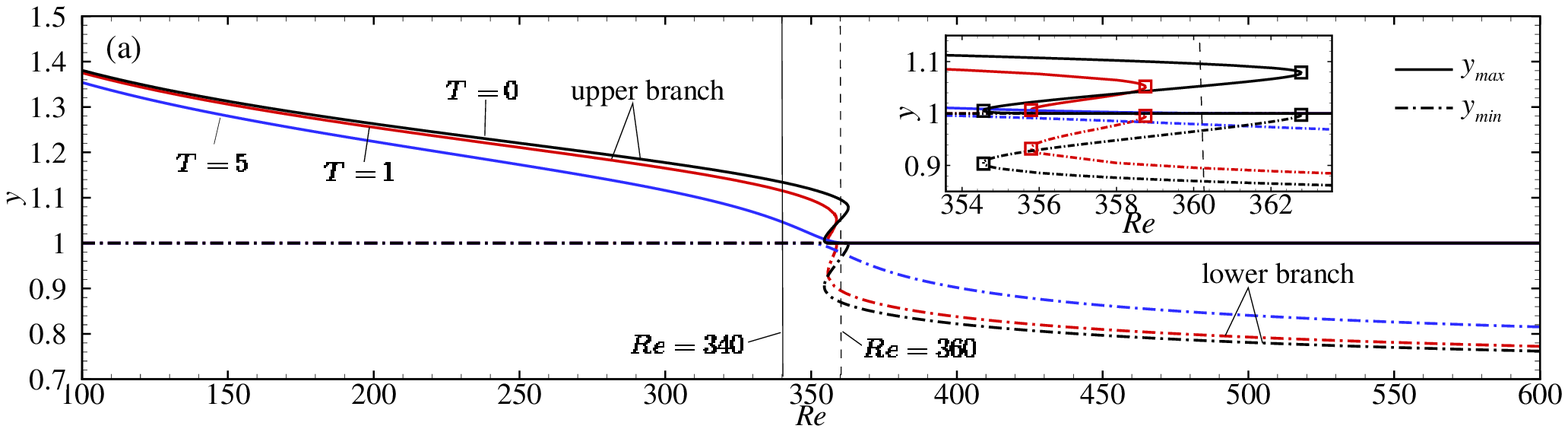}\\
\includegraphics[width=0.49\textwidth]{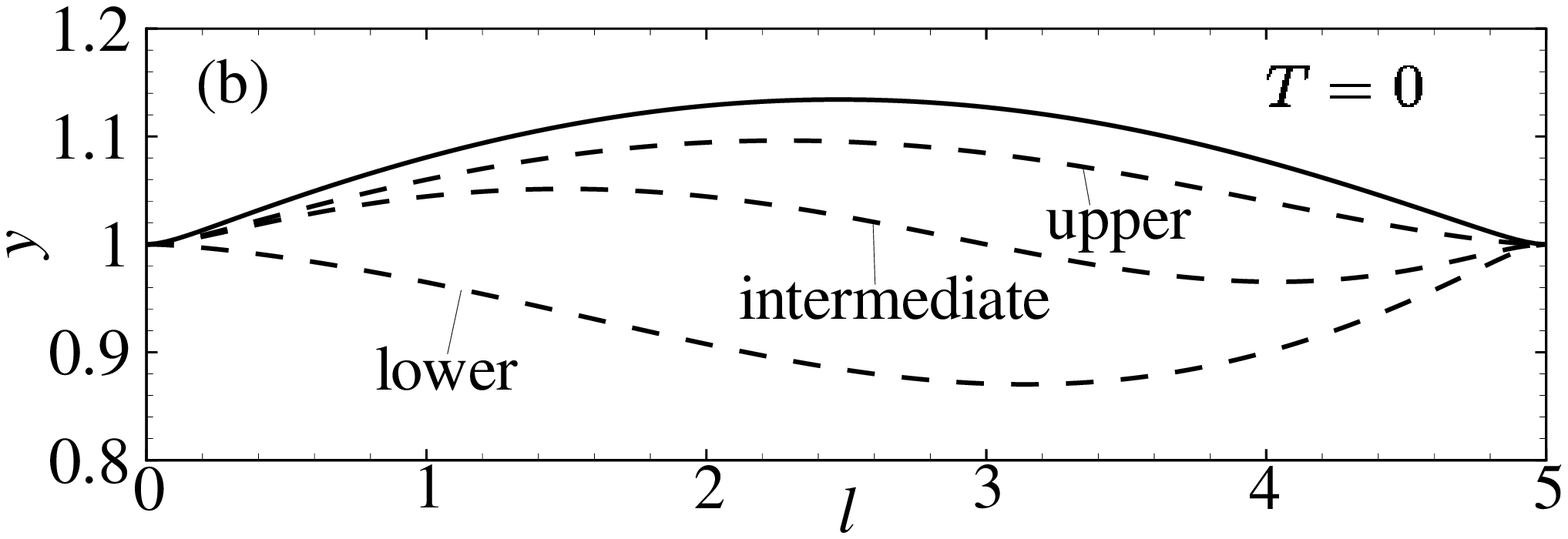}
\includegraphics[width=0.49\textwidth]{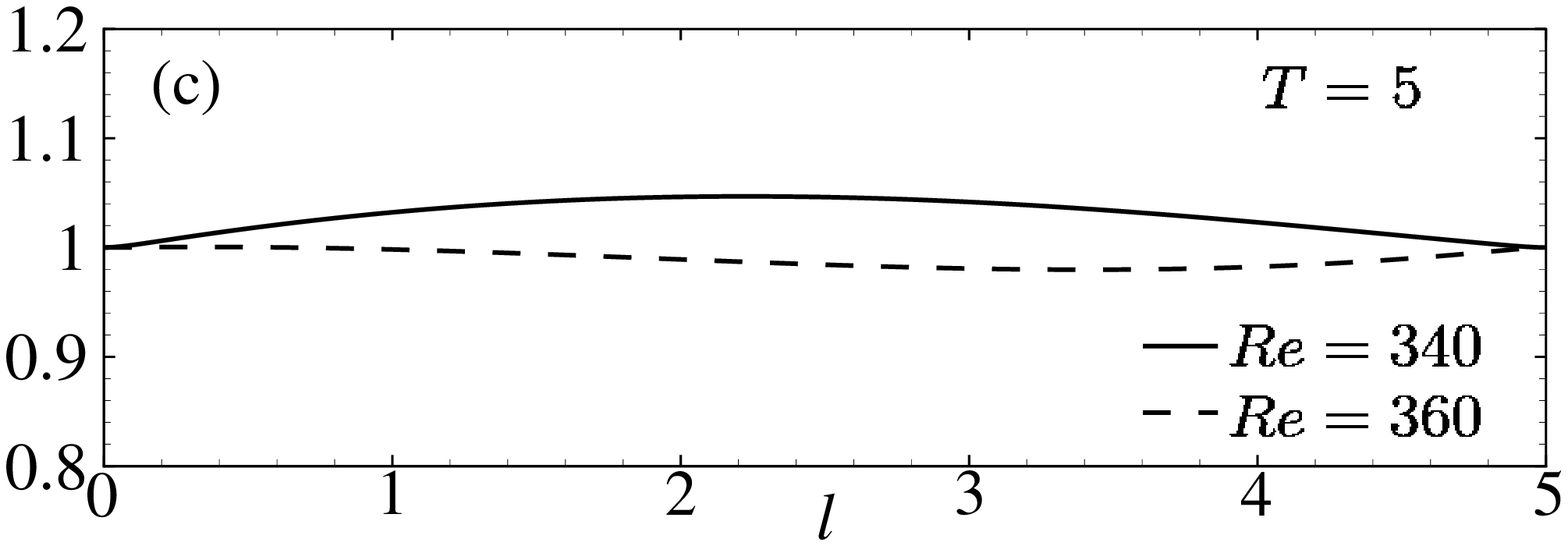}\\
\includegraphics[width=\textwidth]{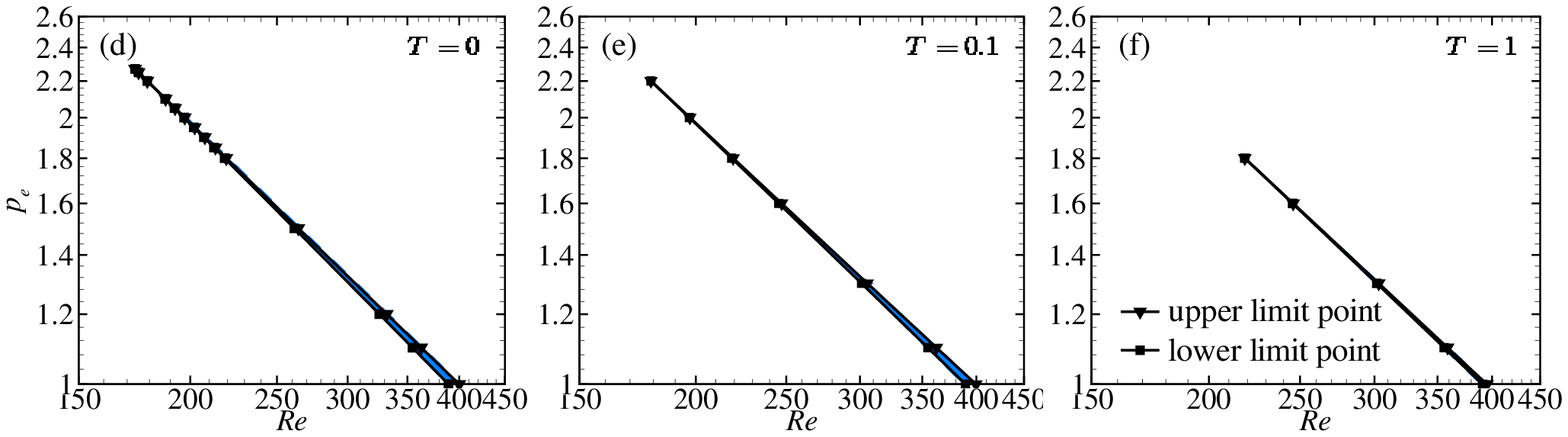}
\caption{Steady solutions for fixed beam thickness $h=0.01$: (a) maximal ($y_{max}$, solid line) and minimal ($y_{min}$, dash-dot line) beam position against $Re$ for $T=0$ (color black), $T=1$ (color red) and $T=5$ (color blue) for $p_e=1.1$;  (b), (c) The steady beam shapes for $p_e=1.1$ with $Re=340$ (solid line) and $Re=360$ (dashed line) for $T=0$ and $T=5$, respectively. (d)-(f) The upper and lower limit points in parameter space spanned by the external pressure $p_e$ and Reynolds number $Re$ for $T=0$, $T=0.1$ and $T=1$, respectively. Note the vertical solid and dashed lines in panel (a) are at $Re=340$ and $360$, respectively. The region with multiple steady solutions is shaded in blue in panels (d)-(f).}
\label{Fig:st_ck0016_fix}
\end{figure}

In order to investigate the role of pre-tension $T$ in setting the steady beam shape, Fig. \ref{Fig:st_ck0016_fix} summarises the steady solutions of the maximal and minimal beam positions (Fig. \ref{Fig:st_ck0016_fix}a), several typical beam shapes (Fig. \ref{Fig:st_ck0016_fix}b) and the upper and lower limit points (Fig. \ref{Fig:st_ck0016_fix}c) with different values of pre-tension with fixed wall thickness $h=0.01$. In particular, the maximal ($y_{max}$, solid line) and minimal ($y_{min}$, dash-dot line) beam deflection on the $y-$direction as a function of Reynolds number $Re$ is shown in Fig. \ref{Fig:st_ck0016_fix}(a) for  $T=0$ (black), $T=1$ (red) and $T=5$ (blue) at fixed external pressure $p_e=1.1$. For low Reynolds numbers the beam is fully inflated (i.e. $y_{min}=1$); this is the upper branch of steady solutions.
As the Reynolds number increases the elastic beam becomes increasingly collapsed though the Bernoulli effect. 
Notably, for $T=0$ the system enters a region with three steady states for $354.56\lesssim Re \lesssim 362.78$ similar to our previous model \citep{wang2021energetics}. As the Reynolds number further increases, the system again exhibits a unique steady state for $Re>362.78$; this branch is the lower branch of steady solutions.
The upper and lower branches of steady solutions are connected by an intermediate branch, which they intersect at the upper and lower limit points, respectively.

A similar region with multiple steady states is observed for $T=1$  ($355.79\lesssim Re \lesssim 358.75$), although the beam pre-tension has significantly narrowed the range of Reynolds numbers for which it is evident for this choice of external pressure.
The region with multiple steady states vanishes entirely for large pre-tension ($T=5$, blue lines in Fig. \ref{Fig:st_ck0016_fix}a).

To illustrate these steady configurations in detail, Fig. \ref{Fig:st_ck0016_fix}(b,c) illustrates the possible steady wall shapes at both $Re=340$ and $Re=360$ for $T=0$ and $T=5$, respectively. At $Re=340$, located on the upper branch for $T=0$ (see the solid vertical line in Fig. \ref{Fig:st_ck0016_fix}a), the steady wall shape is fully inflated and is unique (solid line in Fig. \ref{Fig:st_ck0016_fix}b). Conversely, at $Re=360$, located in the region with multiple steady states for $T=0$ (see the dashed vertical line in Fig. \ref{Fig:st_ck0016_fix}a), there are three possible steady wall shapes (Fig. \ref{Fig:st_ck0016_fix}b).
Whereas for $T=5$ (Fig. \ref{Fig:st_ck0016_fix}c), the system has a unique steady state for both Reynolds numbers and the wall is significantly less deflected due to the larger pre-tension.

Fig. \ref{Fig:st_ck0016_fix}(d-f)  summarises the region with multiple steady states for $T=0$ (Fig. \ref{Fig:st_ck0016_fix}d), $T=0.1$ (Fig. \ref{Fig:st_ck0016_fix}e) and $T=1$ (Fig. \ref{Fig:st_ck0016_fix}f) in parameter space spanned by the external pressure and Reynolds number. In each case we see a triangular region with multiple steady states with external pressure below a critical value and Reynolds number above a critical value. As $T$ increases, this critical point is gradually pushed to larger Reynolds number  and lower external pressures.
However, the region with multiple steady states (Fig. \ref{Fig:st_ck0016_fix}f) remains of approximately the same width relative to the critical point.
In summary, large pre-tension suppresses multiple steady states of the system by shifting the critical point for multiple solutions across the parameter space.

\subsection{Role of Beam Thickness}\label{sec:st-ck}

\begin{figure}[t!]
\centering
\includegraphics[width=\textwidth]{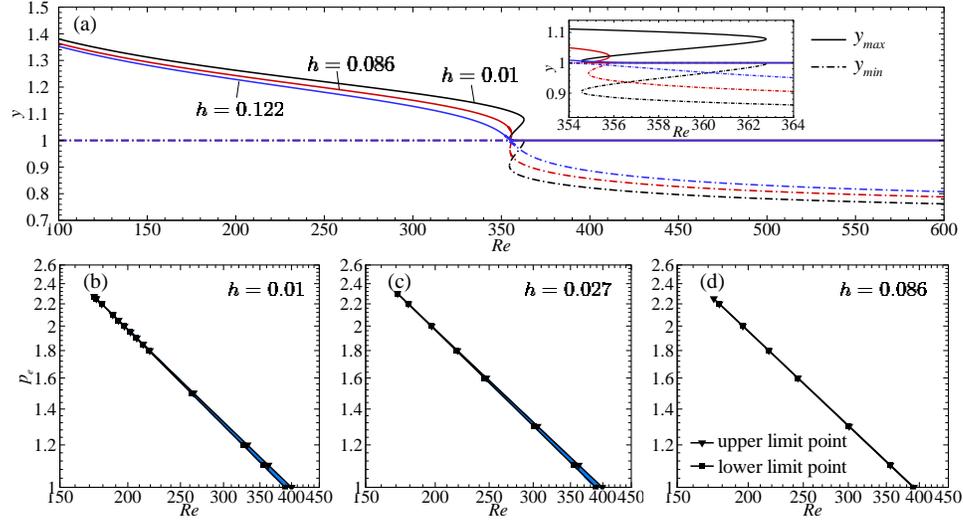}
\caption{Steady solutions for $T=0$, showing (a) the maximal ($y_{max}$, solid line) and minimal ($y_{min}$, dash-dot line) beam position against $Re$ for $h=0.01$ (color black), $h=0.086$ (color red) and $h=0.122$ (color blue); the upper and lower limit points in parameter space spanned by the external pressure $p_e$ and Reynolds number $Re$ for $h=0.01$ (b), $h=0.027$ (c) and $h=0.086$ (d), respectively. In panels (b)-(d), the region with multiple steady solutions is shaded in blue.}
\label{Fig:st_T0_fix}
\end{figure}

To investigate the role of increased bending stiffness on the steady configuration of the beam, Fig. \ref{Fig:st_T0_fix} demonstrates the steady solutions with various beam thickness, plotting the maximal  ($y_{max}$) and minimal ($y_{min}$) beam deflection against the Reynolds number $Re$ for three beam thicknesses ($h=0.01, ~0.086$, and $0.122$) with fixed pre-tension $T=0$ and external pressure $p_e=1.1$ in Fig. \ref{Fig:st_T0_fix}(a). Similar to Fig. \ref{Fig:st_ck0016_fix}, the system shows multiple steady states for $h=0.01$ ($354.56\lesssim Re \lesssim 362.78$); this region of multiple solutions narrows with increased beam thickness $h=0.086$  ($354.87\lesssim Re \lesssim 355.8$). Additionally, the steady system is unique for sufficiently large beam thickness $h=0.122$. 
Fig. \ref{Fig:st_T0_fix}(b-d) summarises the corresponding upper and lower limit points in parameter space spanned by the external pressure $p_e$ and Reynolds number $Re$ for the same three beam thicknesses, showing that the region with multiple steady states narrows as $h$ increases, but the critical point for the existence of multiple solutions does not vary much across the parameters tested. 
In summary, this figure demonstrates that increasing the beam thickness (and hence the bending stiffness) eventually suppresses multiple steady states; however, this is achieved by narrowing the tongue while holding the critical point fixed.

\section{Self-excited Oscillations}\label{sec:unsteady}
In order to test the stability of the system for a given parameter combination we apply a small increment to the steady solution (here we use the steady solution corresponding to a $1\%$ increase in the extensional stiffness $c_{\lambda}$). The system is deemed stable if the unsteady solution converges to its corresponding steady solution, and unstable if the perturbation grows \citep{drazin2002introduction}. The state between stable and unstable is termed as neutrally stable.
In this model the perturbation grows in an oscillatory manner and saturates into a finite amplitude limit cycle.
In this study, we report the dynamics of this oscillatory limit cycle and ignore the initial transient. In particular, we present phase portraits of the oscillation as a function of the wall pressure at the upstream and downstream ends of the compliant segment (e.g.~Fig. \ref{Fig:pmid-pe148-225}c-f, i-l and Fig. \ref{Fig:pmid-st-avg-Tt0-ck0016}b-e) and compute the fluid pressure on the wall at the channel midpoint time-averaged over a period of oscillation (e.g.~Fig. \ref{Fig:pmid-pe148-225}a,g, Fig. \ref{Fig:pmid-st-avg-Tt0-ck0016}a, Fig. \ref{Fig:pmid-st-avg-T0-ck1-2} and Fig. \ref{Fig:pmid-st-avg-T1-5}). In this section, we focus on the unsteady solutions for fixed extensional stiffness ($c_{\lambda}=1600$), testing the stability of both the upper and lower steady branches. In particular, we investigate oscillatory solutions with no pre-tension and low beam thickness (Sec. \ref{sec:un_lowbending}), large beam thickness (Sec. \ref{sec:un_largebending}) and large pre-tension (Sec. \ref{sec:un_largeT}).

\subsection{Multiple Oscillatory Solutions with Low Bending Stiffness}\label{sec:un_lowbending}

\begin{figure}[t!]
\centering
\includegraphics[width=0.6\textwidth]{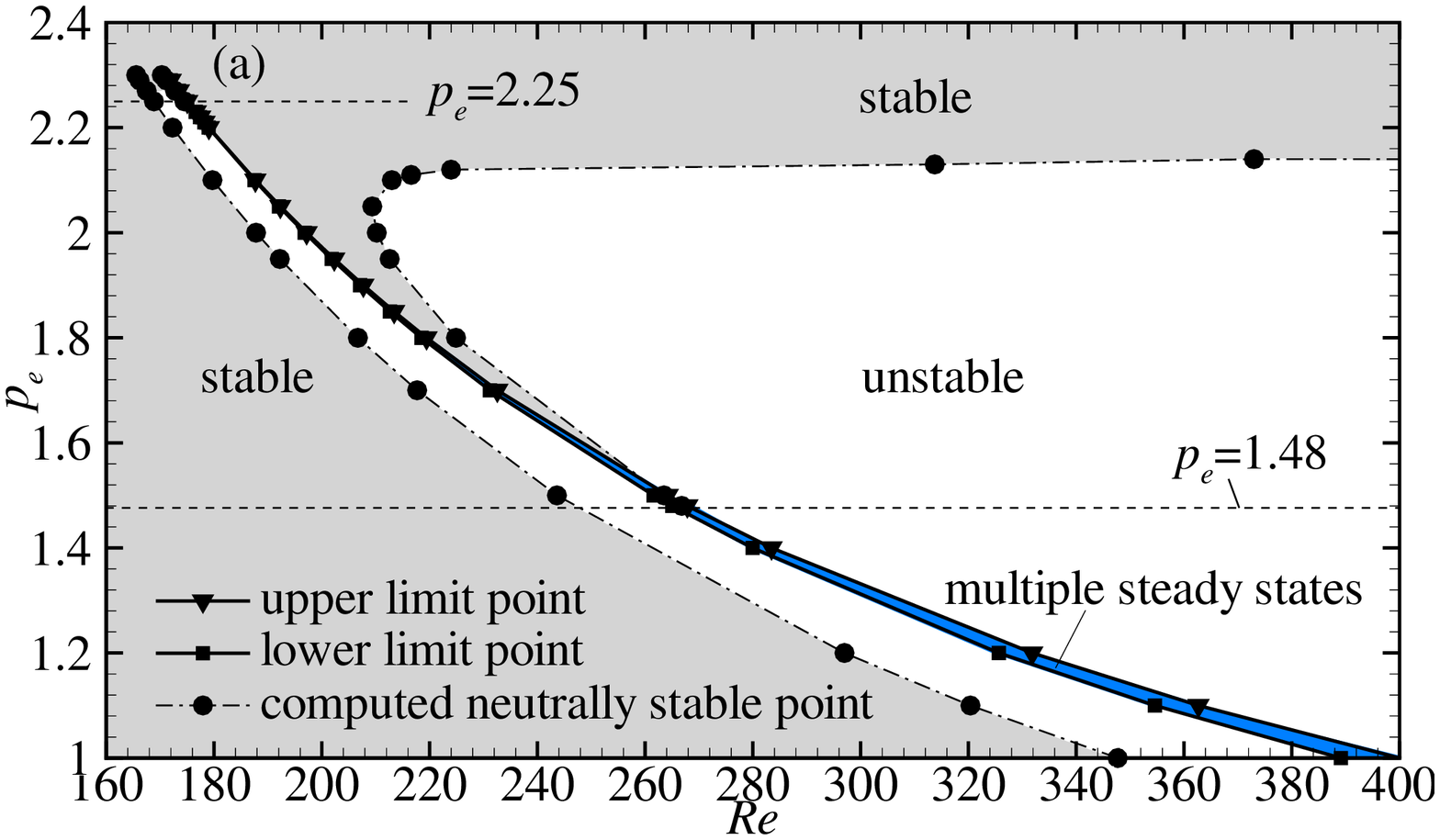}
\includegraphics[width=0.3\textwidth]{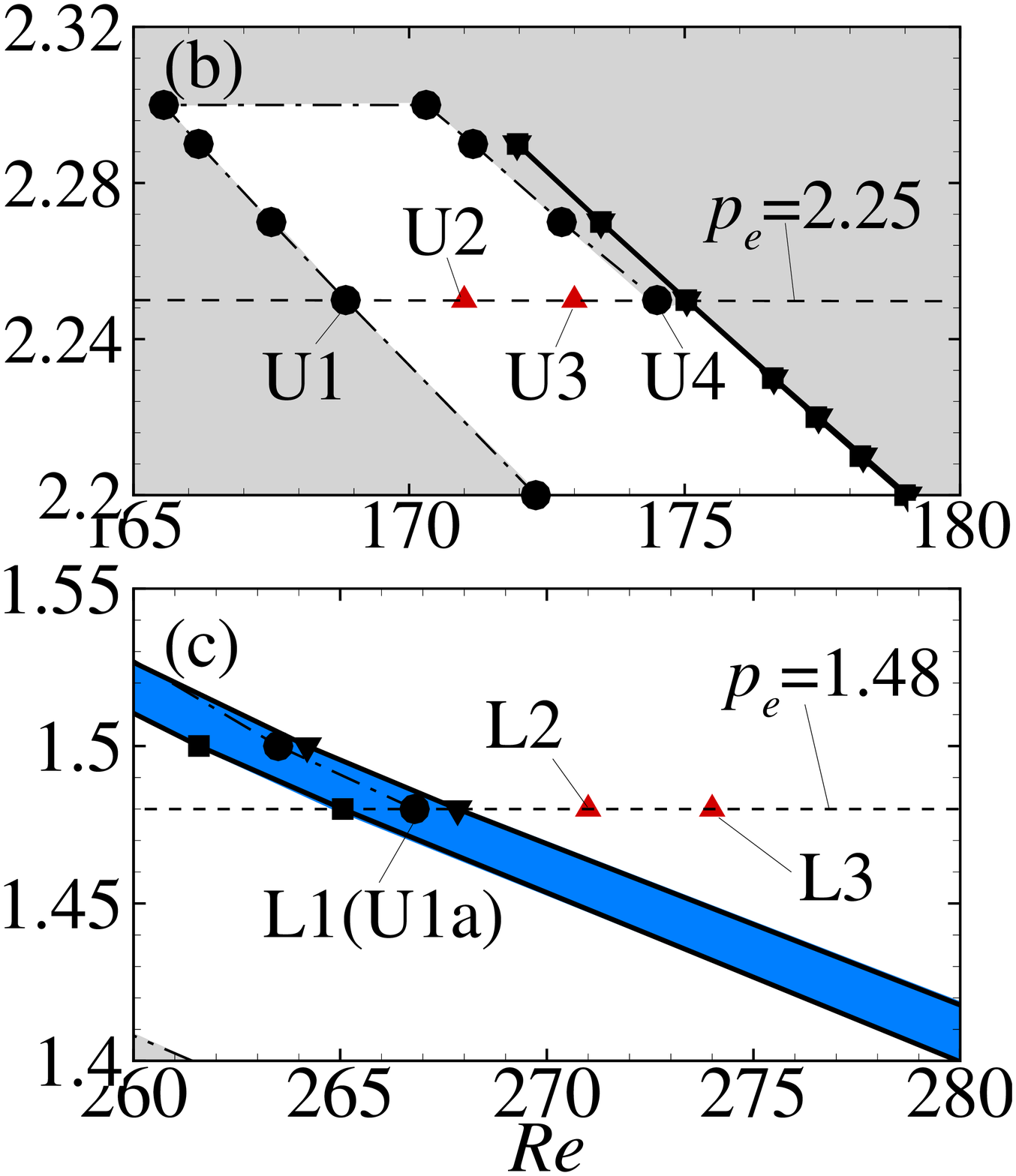}
\caption{An overview of the parameter space spanned by the Reynolds number and external pressure at $T=0$ and $h=0.01$. The computed neutrally stable points are denoted as filled black circles, and the upper and lower limit points are denoted as filled black squares connected with dashed and solid lines. The region where the system is stable is shaded in grey and the region with multiple steady solutions is shaded in blue. (b), (c) show the zoom-in of the region near $p_e=2.25$ and $p_e=1.48$, respectively. Note the unsteady solutions of the operating points for $p_e=2.25$: U1 ($Re=168.85$), U2 ($Re=171$), U3 ($Re=173$), U4 ($Re=174.5$); and for $p_e=1.48$: L1 (U1a) ($Re=266.8$), L2 ($Re=271$), L3 ($Re=274$) are shown in Fig. \ref{Fig:pmid-pe148-225}.}
\label{Fig:pe_re_t0-ck0016}
\end{figure}

\begin{figure}
\centering
\includegraphics[width=0.49\textwidth]{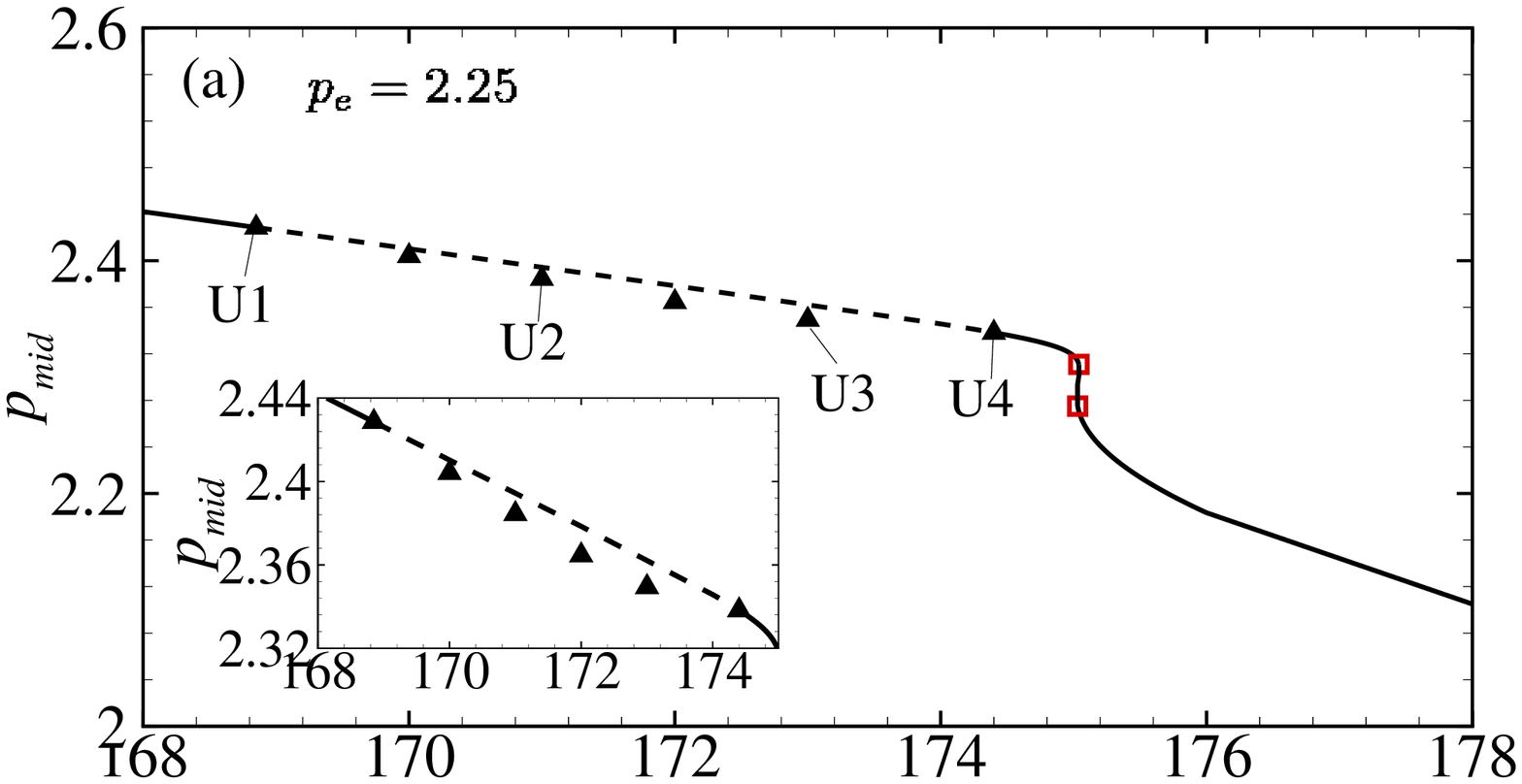}
\includegraphics[width=0.49\textwidth]{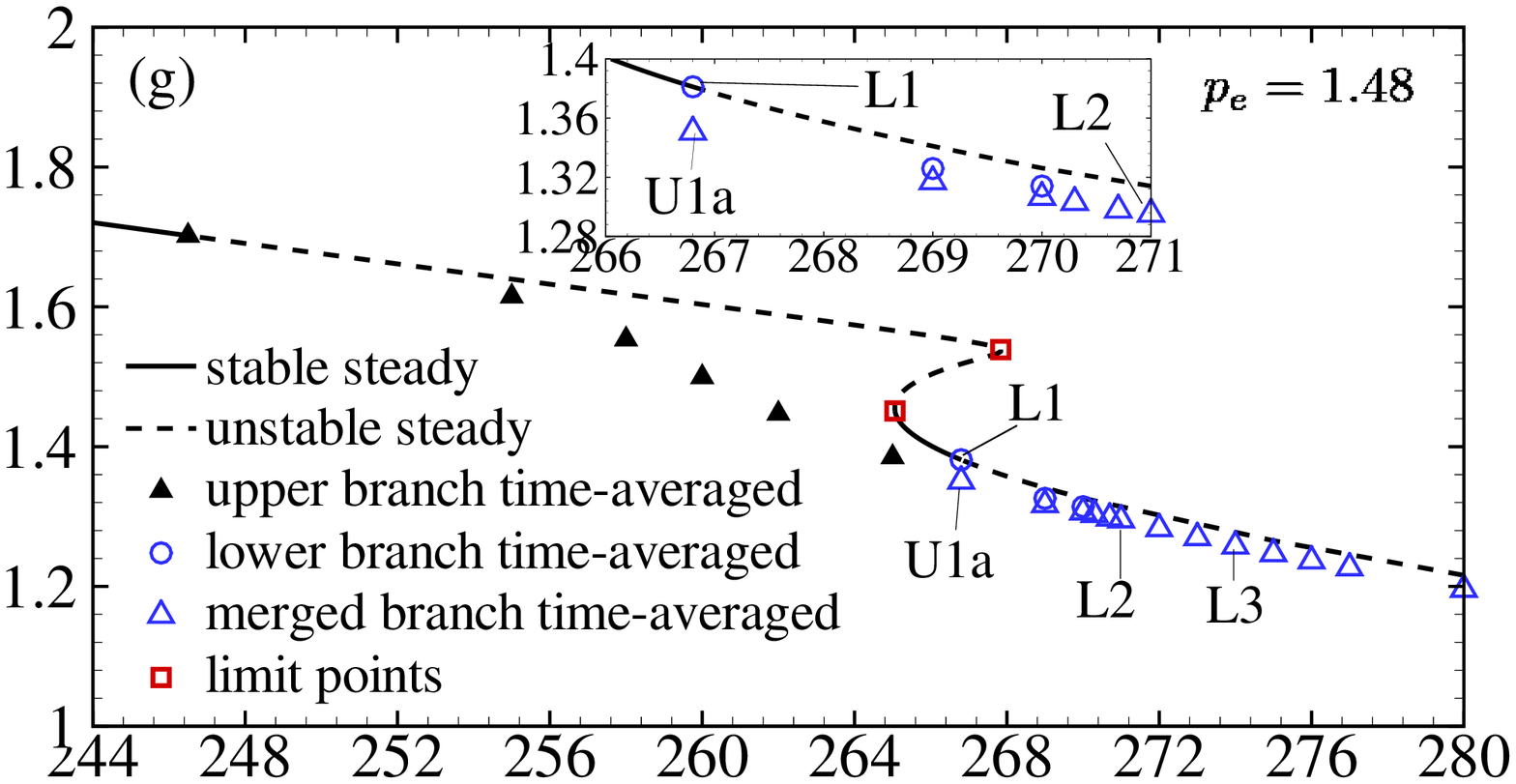}\\
\includegraphics[width=0.49\textwidth]{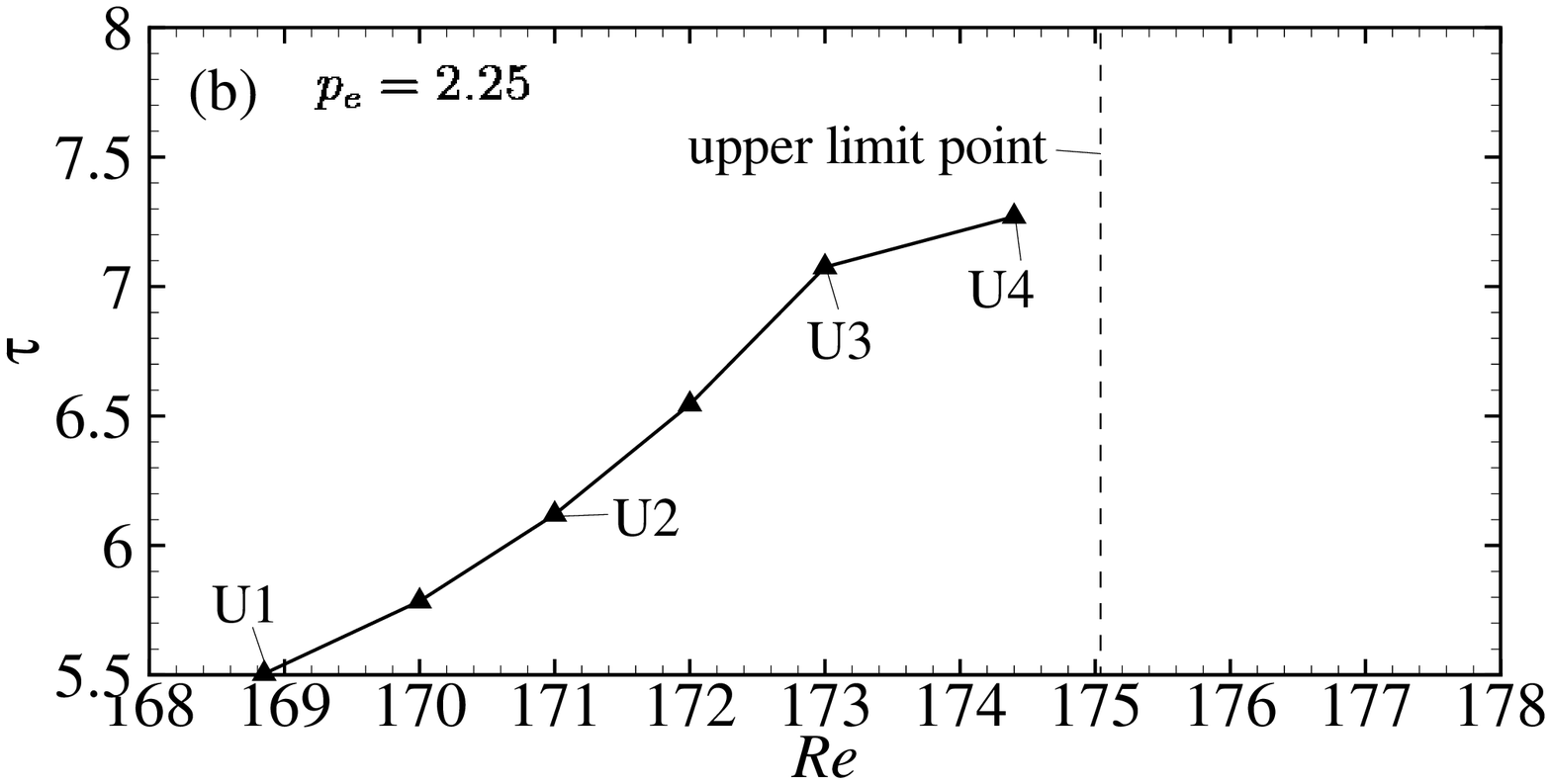}
\includegraphics[width=0.49\textwidth]{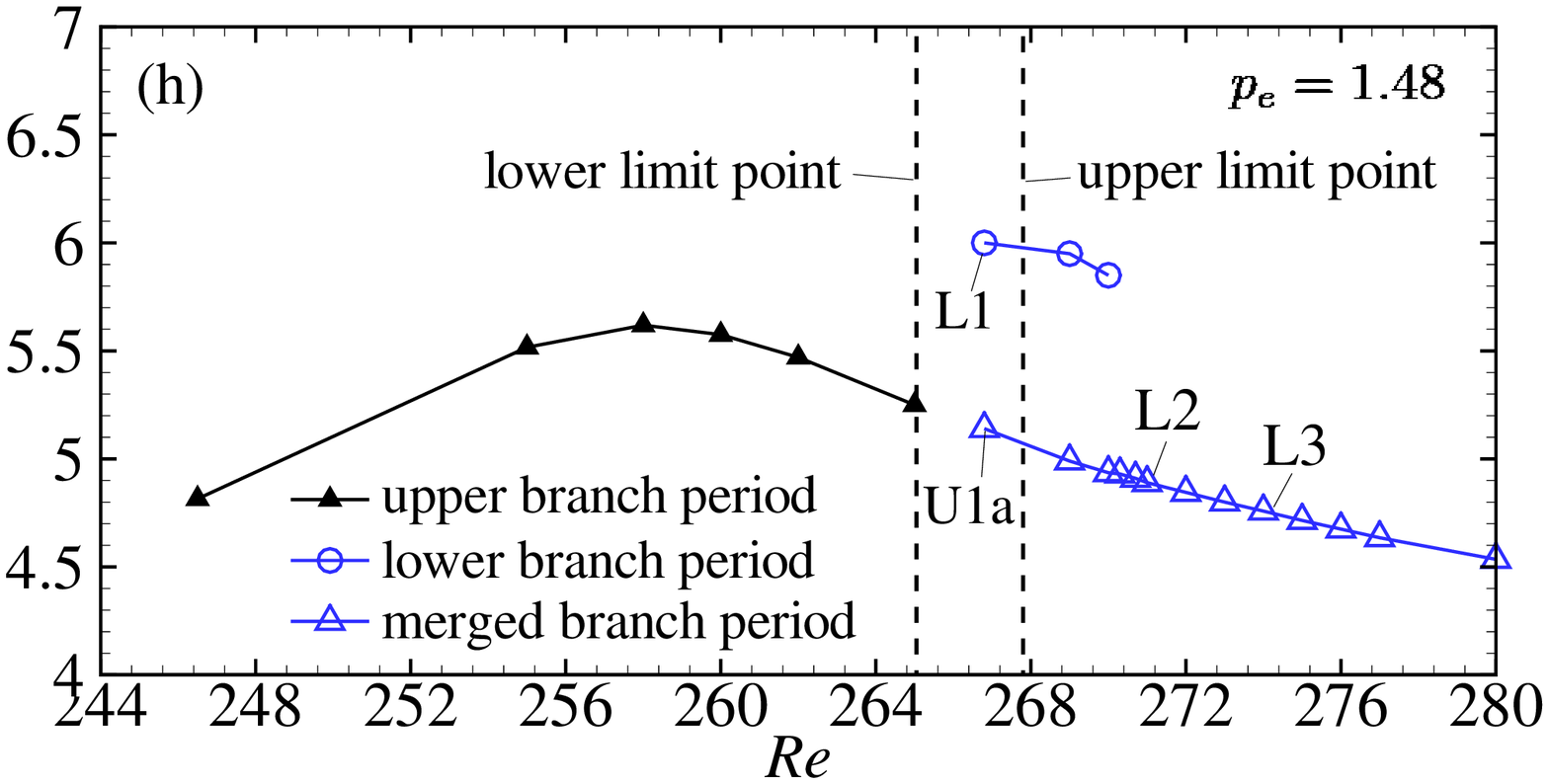}\\
\includegraphics[width=0.49\textwidth]{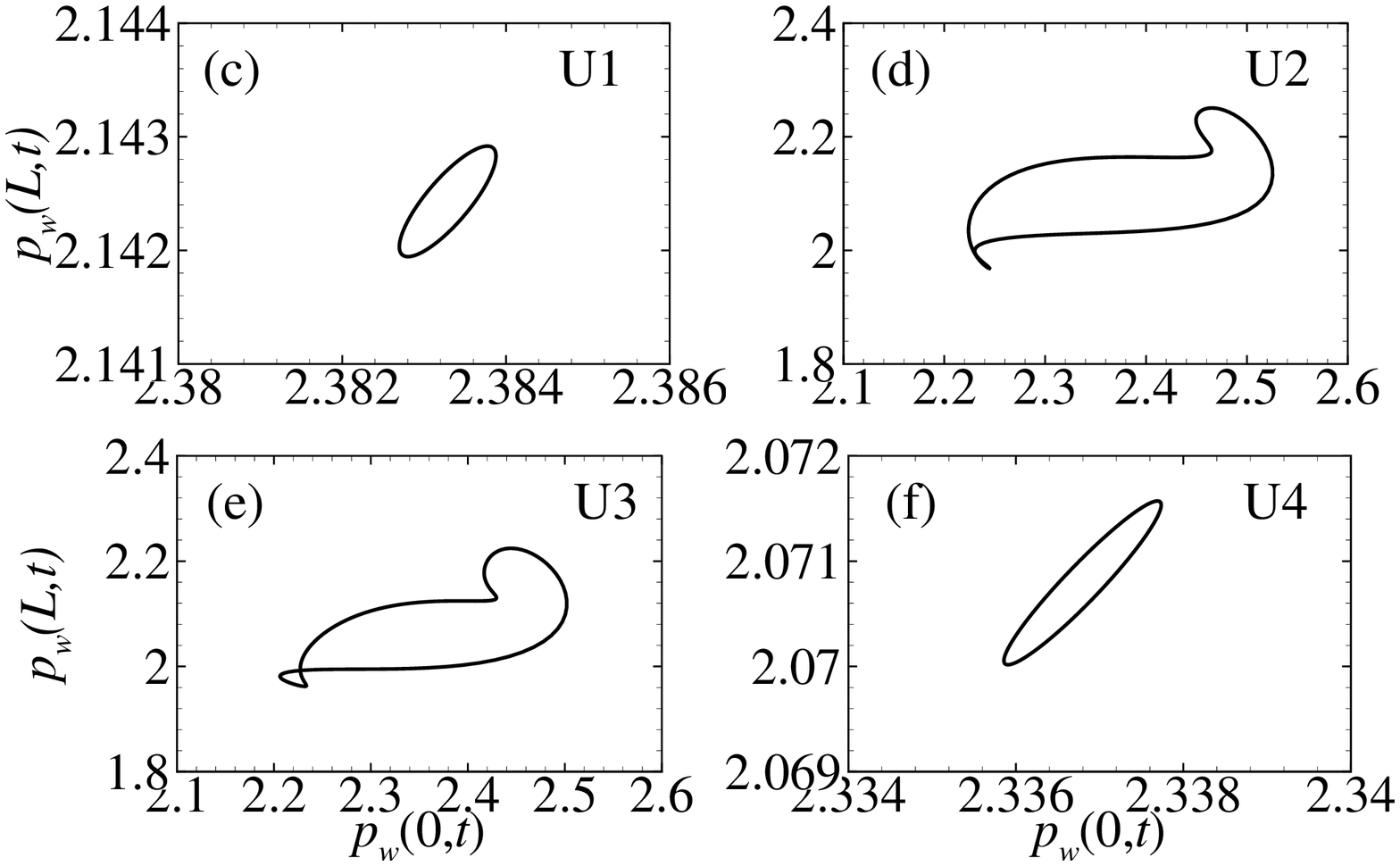}
\includegraphics[width=0.49\textwidth]{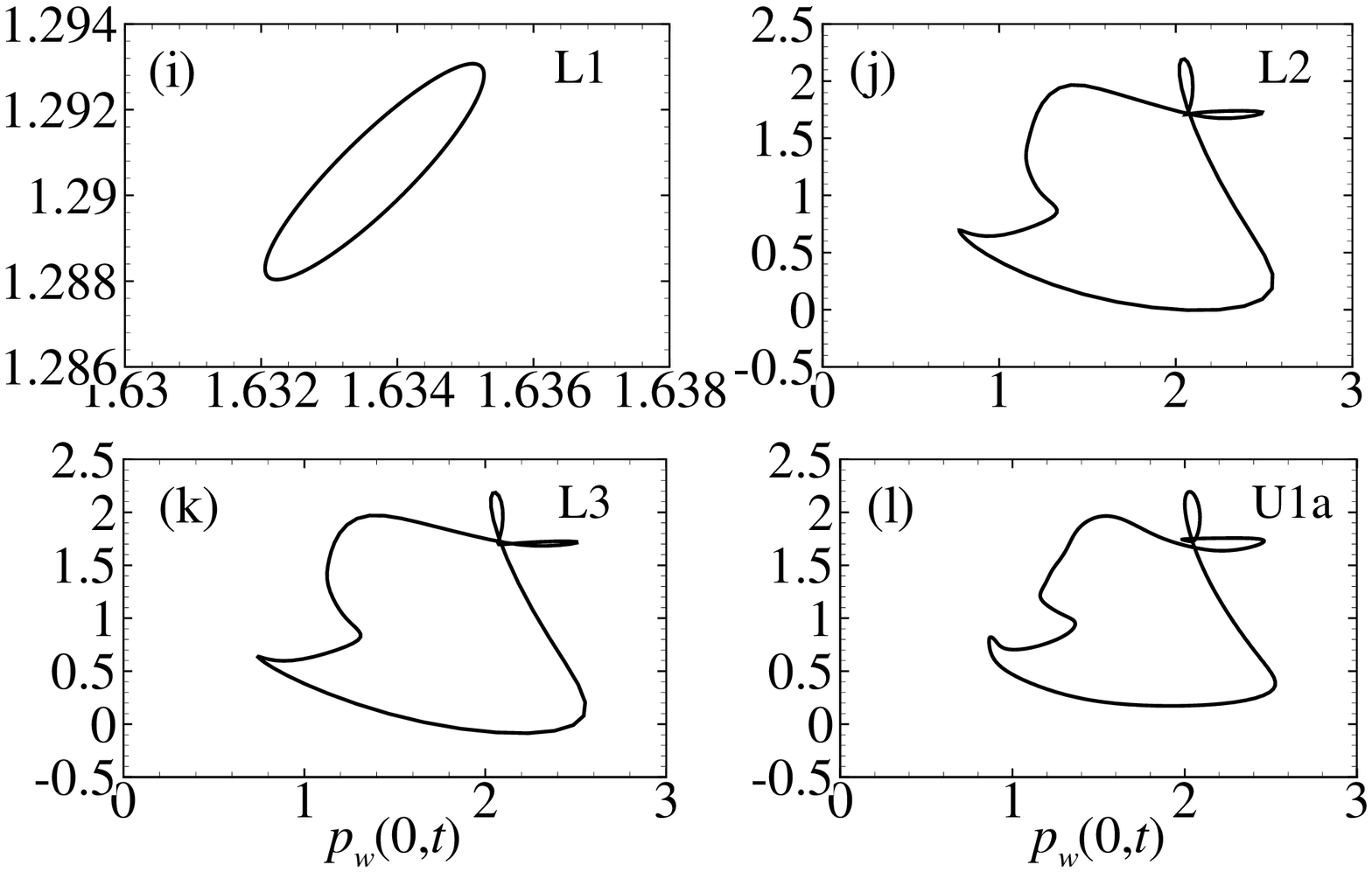}
\caption{Unsteady solutions of the upper and lower steady branches for $p_e=2.25$ and $p_e=1.48$. (a, g) Bifurcation diagram of midpoint wall pressure as a function of the Reynolds number for $p_e=2.25$ and $p_e=1.48$, respectively; (b, h) Oscillation period as a function of the Reynolds number for $p_e=2.25$ and $p_e=1.48$, respectively; (c-f, i-l) Phase portraits in the space spanned by the wall pressure measured at the upstream and downstream ends of the compliant segment for $p_e=2.25$ and $p_e=1.48$, respectively.}
\label{Fig:pmid-pe148-225}
\end{figure}

\begin{figure}[t!]
\centering
\includegraphics[width=0.48\textwidth]{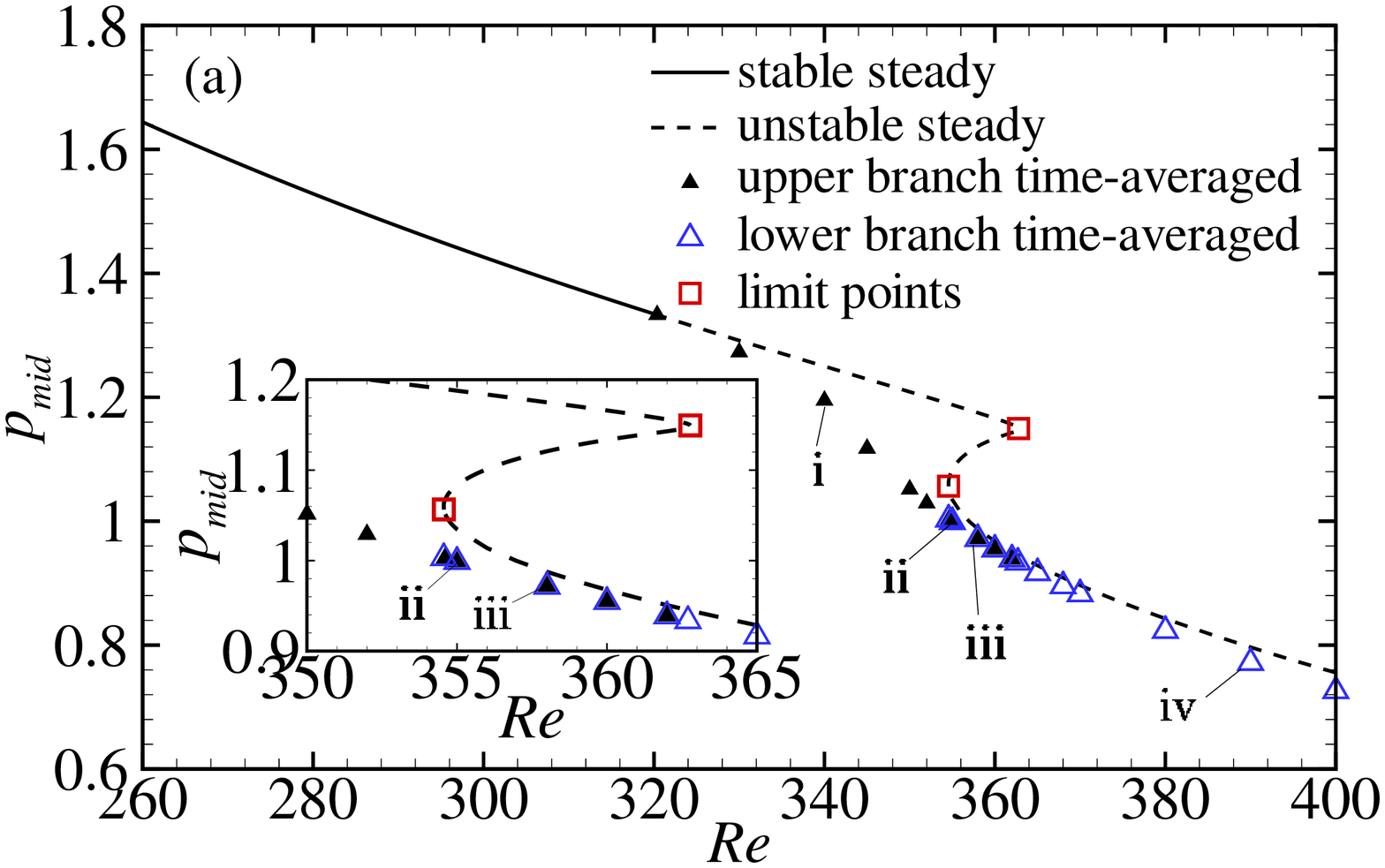}
\includegraphics[width=0.48\textwidth]{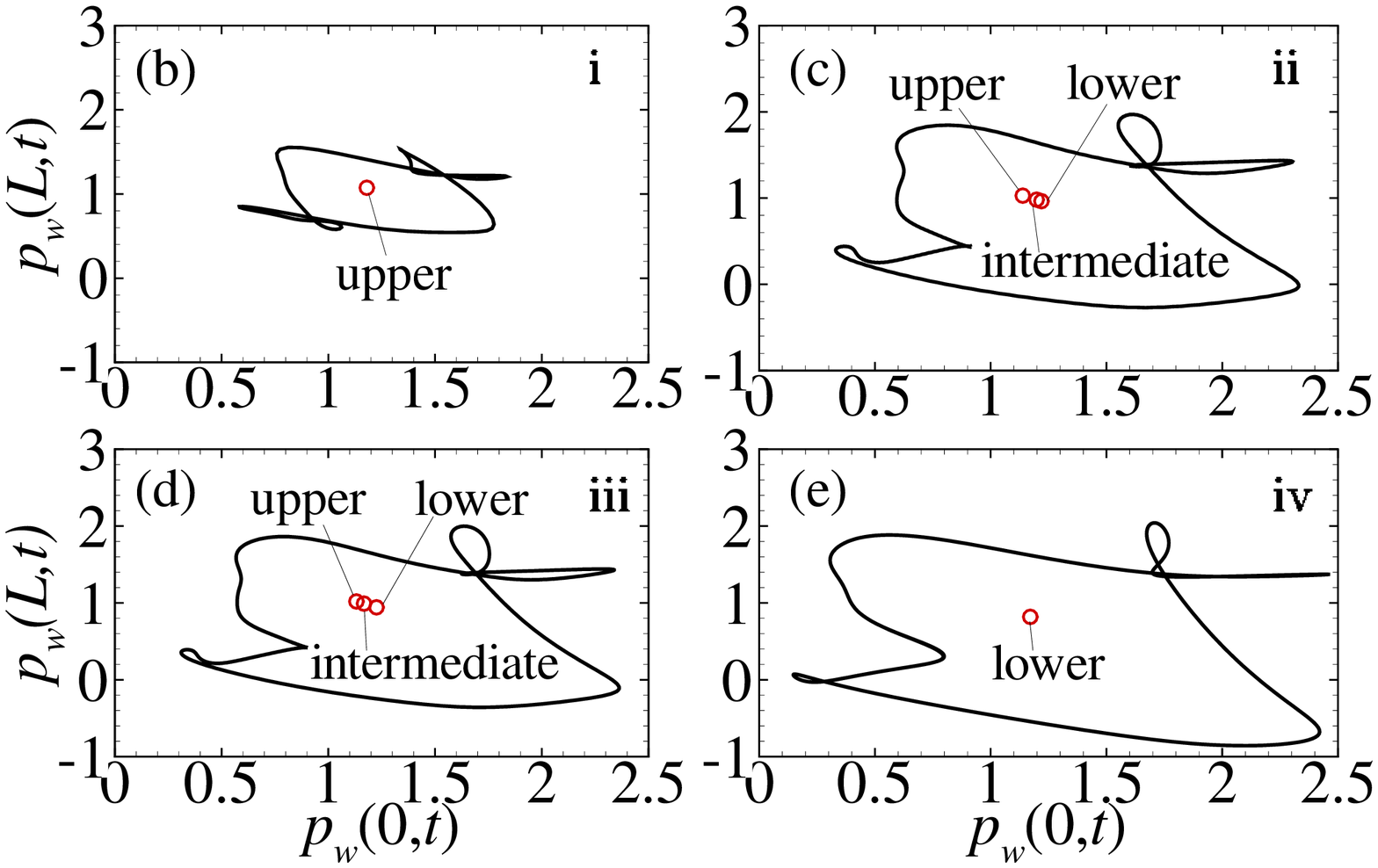}
\caption{Unsteady bifurcation diagram of the system for $T=0$, $h=0.01$ and $p_e=1.1$, showing (a) The midpoint wall pressure $p_{mid}$ as a function of the Reynolds number $Re$. The steady midpoint pressure $p_{mid}$ (solid (stable) and dashed (unstable) lines), the time-averaged midpoint wall pressure $p_{mid}^{avg}$ (triangles), the upper and lower branch limit points (squares); (b-e) Phase portraits in space spanned by the wall pressure measured at the upstream and downstream ends of the compliant segment at operating points {\romannumeral 1}-{\romannumeral 4} labelled in (a). The corresponding values of the upper, intermediate and lower steady branches are denoted by open circles.}
\label{Fig:pmid-st-avg-Tt0-ck0016}
\end{figure}

\begin{figure}[t!]
\centering
\includegraphics[width=0.95\textwidth]{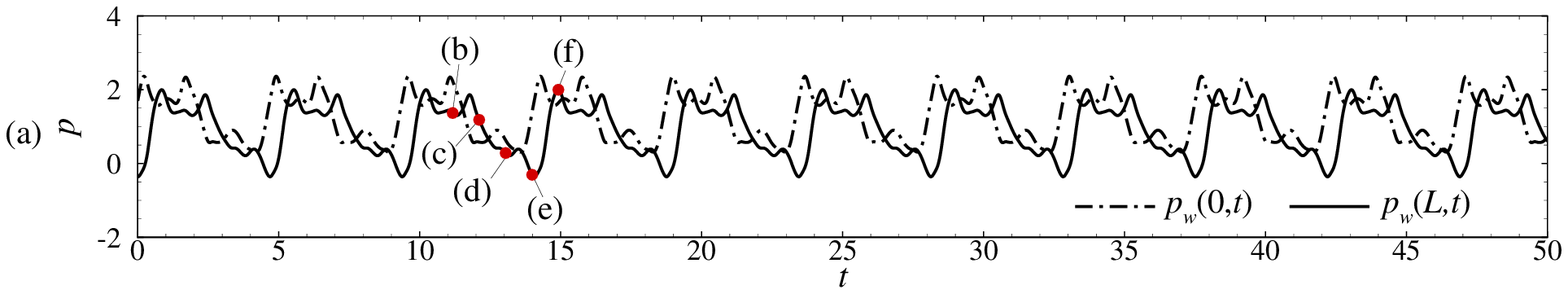}\\
\includegraphics[width=0.95\textwidth]{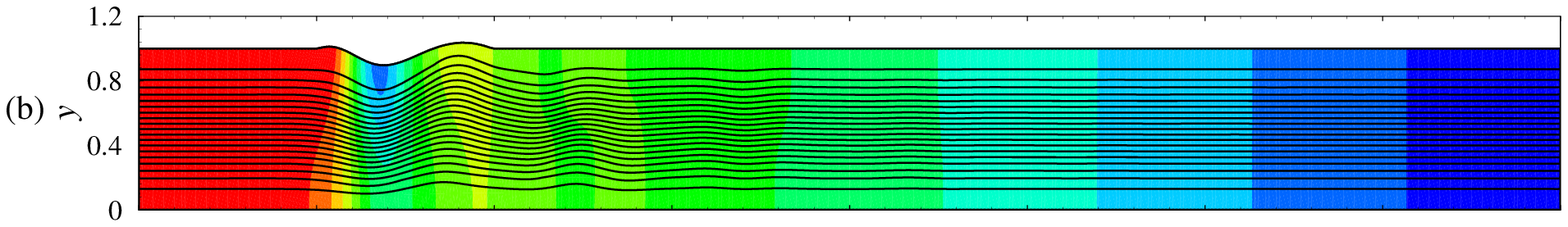}\\
\includegraphics[width=0.95\textwidth]{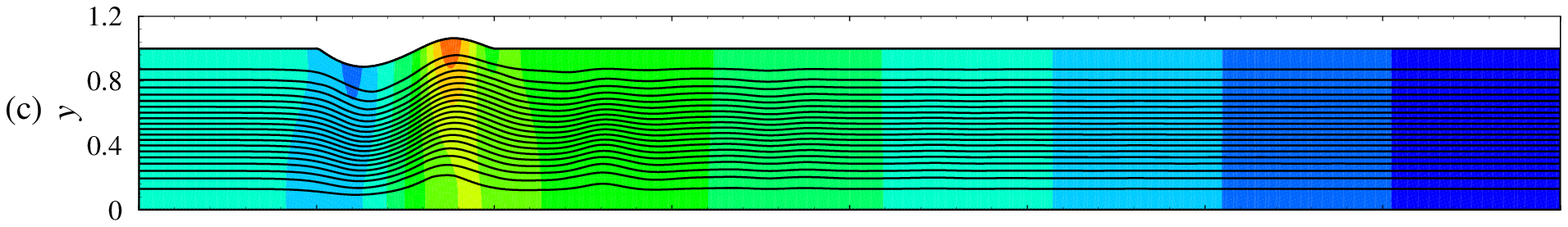}\\
\includegraphics[width=0.95\textwidth]{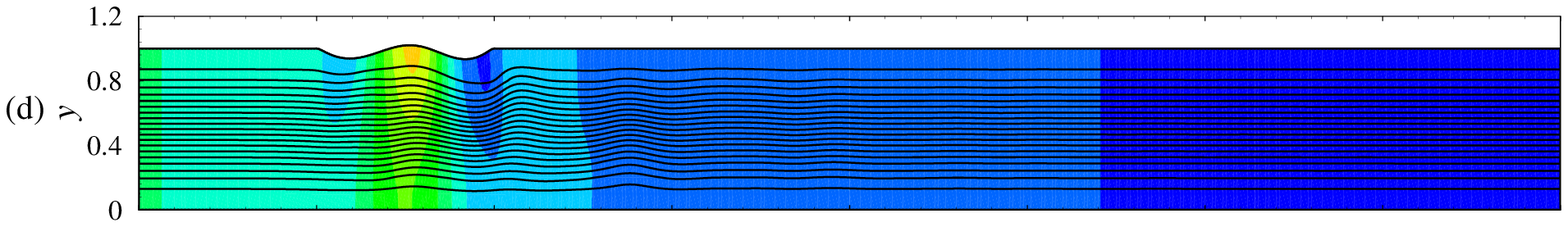}\\
\includegraphics[width=0.95\textwidth]{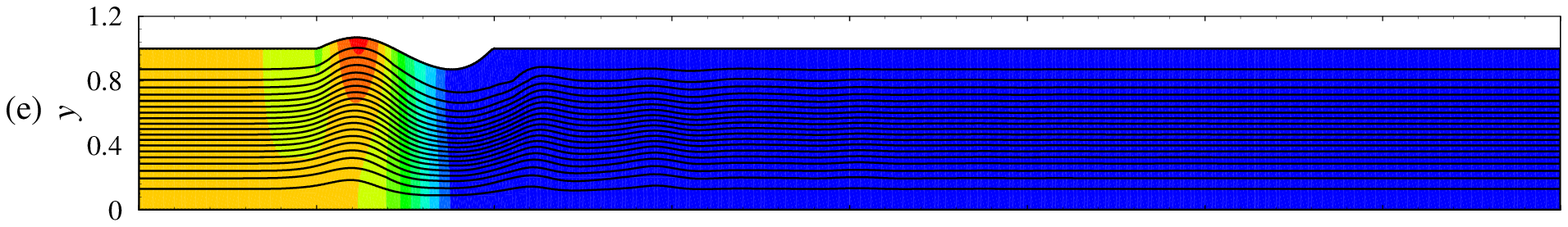}\\
\includegraphics[width=0.95\textwidth]{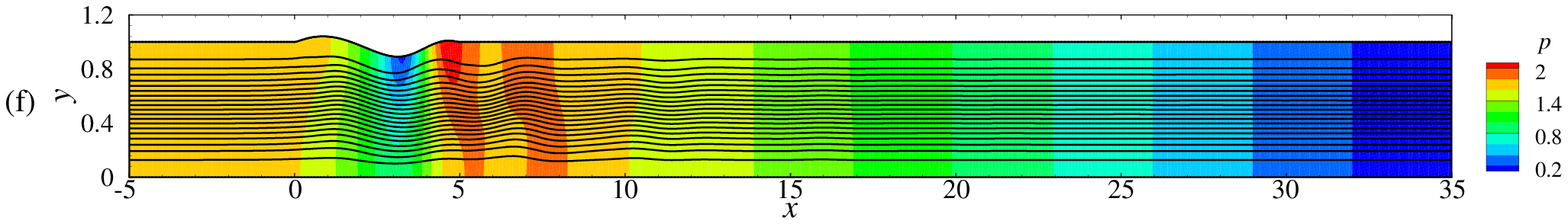}
\caption{Self-excited oscillations arising at operating point {\romannumeral 3} ($Re=358$, $p_e=1.1$, $T=0$ and $h=0.01$ labelled in Fig. \ref{Fig:pmid-st-avg-Tt0-ck0016}a), showing (a) time history of the wall pressure measured at upstream ($p_w(0,t)$) and downstream ($p_w(L,t)$) ends of the compliant segment; (b-f) streamlines and pressure contours of the flow at five equally-spaced time instances over a period of oscillation, labelled in (a).}
\label{Fig:pwall-fullfield}
\end{figure}

We first analyse unsteady simulations of perturbations to the upper and lower steady branches with no pre-tension ($T=0$) and a thin beam ($h=0.01$).  
Here we extend our previous analysis \citep{wang2021energetics} to examine the parameter space spanned by the external pressure and Reynolds numbers, as shown in Fig. \ref{Fig:pe_re_t0-ck0016} (Note that in \cite{wang2021energetics} we held the external pressure $p_e=1.95$ throughout). The steady behaviour of the system is similar to the cases presented in Fig. \ref{Fig:st_ck0016_fix}(d-f) and \ref{Fig:st_T0_fix}(b-d), where three steady states can co-exist in a narrow tongue between the upper and lower limit points (blue shaded region marked by filled symbols).
Testing the stability of these steady solutions we find that both the upper and lower steady branches can independently become unstable to oscillations (each via a supercritical Hopf bifurcation) in the neighbourhood of the region of parameter space with multiple steady states.
The computed neutrally stable points from the upper and lower steady branches are marked as filled circles, and the neutral stability curves are denoted as dot-dashed lines connecting these neutral points. 
In this parameter space the upper branch is unstable within a narrow tongue to the left of the trace of the upper branch limit points.
This upper branch neutral curve intersects the trace of the upper branch limit points at a co-dimension 2 (Takens-Bogdanov) point \citep{glendinning1994stability,strogatz2018nonlinear} at $p_e\approx2.25$, $Re\approx174.5$ (we term this the upper Takens-Bogdanov point).
For external pressures less than this upper Takens-Bogdanov point the upper branch oscillation is eventually restabilised as the Reynolds number increases via an interaction between the oscillatory limit cycle and  the upper branch limit point \citep[see details in][]{wang2021energetics}. However, this unstable tongue extends to slightly larger external pressures than those which admit see multiple steady states. 
For external pressures greater than the upper Takens-Bogdanov point the upper branch oscillation instead restabilises via a second Hopf bifurcation as the Reynolds number increases. This interaction is explored further in Fig. \ref{Fig:pmid-pe148-225}, where we examine the time-averaged midpoint pressure of fully developed upper branch oscillations (Fig. \ref{Fig:pmid-pe148-225}a), their corresponding period (Fig. \ref{Fig:pmid-pe148-225}b) and illustrate several limit cycles of oscillation (Fig. \ref{Fig:pmid-pe148-225}c-f). As in \cite{wang2021energetics}, this upper branch oscillation is associated with an overall decrease in the time-averaged midpoint pressure compared to the steady state (Fig. \ref{Fig:pmid-pe148-225}a), and the period of oscillation increases as the upper limit point is approached (Fig. \ref{Fig:pmid-pe148-225}b). The oscillation develops a rather complicated limit cycle (Fig. \ref{Fig:pmid-pe148-225}c-f), similar to the upper branch oscillations reported in \cite{wang2021energetics}. 

Furthermore, the lower branch of steady solutions is unstable within a tongue to the right of the region with multiple steady states. This tongue is stabilised for large external pressure ($p_e \gtrsim 2.12$), and tracks close to the curve of lower branch limit points, merging at a second (lower) Takens-Bogdanov point at $Re\approx266.8$, $p_e\approx1.48$. The neighbourhood of this lower Takens-Bogdanov point is explored in Fig. \ref{Fig:pmid-pe148-225}(g-l), plotting the time-averaged midpoint pressure (Fig. \ref{Fig:pmid-pe148-225}g), period of oscillation (Fig. \ref{Fig:pmid-pe148-225}h) and several limit cycles (Fig. \ref{Fig:pmid-pe148-225}i-l) for fixed external pressure $p_e=1.48$. During oscillation the midpoint pressure is again less than the corresponding steady configuration. 
Just beyond the critical Reynolds number for lower branch instability, the oscillation grows and saturates into an elliptical limit cycle of larger period than the upper branch oscillations (open circles in Fig. \ref{Fig:pmid-pe148-225}h) but of much smaller amplitude (open circles in Fig. \ref{Fig:pmid-pe148-225}g). A phase portrait for point L1 is shown in Fig. \ref{Fig:pmid-pe148-225}(i). For Reynolds numbers sufficiently close to critical ($Re\gtrsim 266.8$) this small-amplitude limit cycle is maintained over the lifetime of our numerical simulations. However, for larger Reynolds numbers ($Re\gtrsim 270.3$) the system visits this limit cycle only transiently and eventually saturates into a limit cycle of larger amplitude (open triangles in Fig. \ref{Fig:pmid-pe148-225}g) and shorter period (open triangles in Fig. \ref{Fig:pmid-pe148-225}h). These nonlinear oscillations are a direct continuation of the oscillations bifurcating from the upper branch steady state, displaying analogous phase portraits (Fig. \ref{Fig:pmid-pe148-225}j,k). Furthermore, across the range of Reynolds numbers which exhibit the small-amplitude limit cycle growing from the lower steady branch ($266.8\lesssim Re \lesssim 270$), it turns out that the system also exhibits a saturated limit cycle analogous to the upper branch oscillations for these points as well (amplitude and period shown as open triangles in Fig. \ref{Fig:pmid-pe148-225}(g,h) and a typical phase portrait is shown in Fig. \ref{Fig:pmid-pe148-225}(l). Hence, this system appears to exhibit hysteresis in the oscillatory behaviour across a small range of parameters with two possible branches of fully developed oscillations (open circles and open triangles in Fig. \ref{Fig:pmid-pe148-225}h); these two branches merge together at $Re\approx270.3$ and then continue to large Reynolds numbers (open triangles in Fig. \ref{Fig:pmid-pe148-225}h). There are two possibilities to explain this observation: either the system exhibits two co-existing limit cycles across a range of parameters, or these reported lower branch (small-amplitude) limit cycles are in fact long transients which eventually grow and saturate to a limit cycle analogous to the upper branch oscillations (note that we see no evidence of this transition for simulations close to the critical Reynolds number over the long time intervals considered in our simulations). Either way, for large enough Reynolds numbers the system exhibits a merged family of oscillations (open triangles in Fig. \ref{Fig:pmid-pe148-225}h) which grow from the lower steady branch but are a direct continuation of nonlinear oscillations bifurcating from the upper steady branch.

It emerges that for external pressures less than the lower Takens-Bogdanov point, this merged family of oscillations becomes more prominent with a typical example for $p_e=1.1$ shown in Fig. \ref{Fig:pmid-st-avg-Tt0-ck0016}, showing the bifurcation diagram (Fig. \ref{Fig:pmid-st-avg-Tt0-ck0016}a) and four typical limit cycles in the neighbourhood of the region with multiple steady states (Fig. \ref{Fig:pmid-st-avg-Tt0-ck0016}b-e). As the Reynolds number increases the system becomes unstable along the upper branch of steady solutions, exhibiting a decrease in the time-averaged midpoint pressure with an elaborate limit cycle around the upper branch steady state (Fig. \ref{Fig:pmid-st-avg-Tt0-ck0016}b). As the Reynolds number increases into the region with multiple steady states the oscillation encompass all three steady solutions (Fig. \ref{Fig:pmid-st-avg-Tt0-ck0016}c, d), and there is now no interaction with the intermediate steady state (a prominent feature of purely upper branch oscillations, as discussed in \cite{wang2021energetics}).
As the Reynolds number continues to increase into the region with only a lower steady state, the oscillation exhibits a qualitatively similar limit cycle (Fig. \ref{Fig:pmid-st-avg-Tt0-ck0016}e).

To further explore the nature of these merged oscillations, Fig. \ref{Fig:pwall-fullfield} examines the oscillation in more detail at a point in parameter space which exhibits multiple steady states, showing time-trace of the wall pressure at the two ends of the compliant segment (Fig. \ref{Fig:pwall-fullfield}a) and several snapshots of the flow-field and pressure contours in the channel (Fig. \ref{Fig:pwall-fullfield}b-e).
The fully developed (nonlinear) oscillation retains many of the characteristics of the upper and lower branch oscillations from which it has merged (these individual oscillations are discussed in \cite{wang2021energetics}). Similar to the upper branch oscillations, the limit cycle involves an upstream propagating hump in the wall profile which is suppressed by interaction with the upstream rigid segment and replaced by a new upstream propagating hump originating at the downstream end of the compliant segment. However, like the lower branch oscillations, the wall profile exhibits significant indentation into the channel (and so much lower fluid pressures). This greater wall indentation leads to the formation of a low pressure region near the end of the compliant segment (Fig. \ref{Fig:pwall-fullfield}d), which creates a (weak) vorticity wave propagating along the downstream rigid segment (Fig. \ref{Fig:pwall-fullfield}e, f), similar to oscillations about a collapsed steady state \citep{luo1996numerical}. Interestingly, the wall pressures at the two ends of the compliant segment are much closer to being in phase than the upper or lower branch oscillations alone. In summary, the figure illustrates the flow profile of our new family of self-excited oscillations, showing that it retains characteristics of both the upper and lower oscillations that have been reported elsewhere \citep{wang2021energetics}.

\subsection{Oscillations with Large Beam Thickness}\label{sec:un_largebending}
\begin{figure}[t!]
\centering
\includegraphics[width=0.48\textwidth]{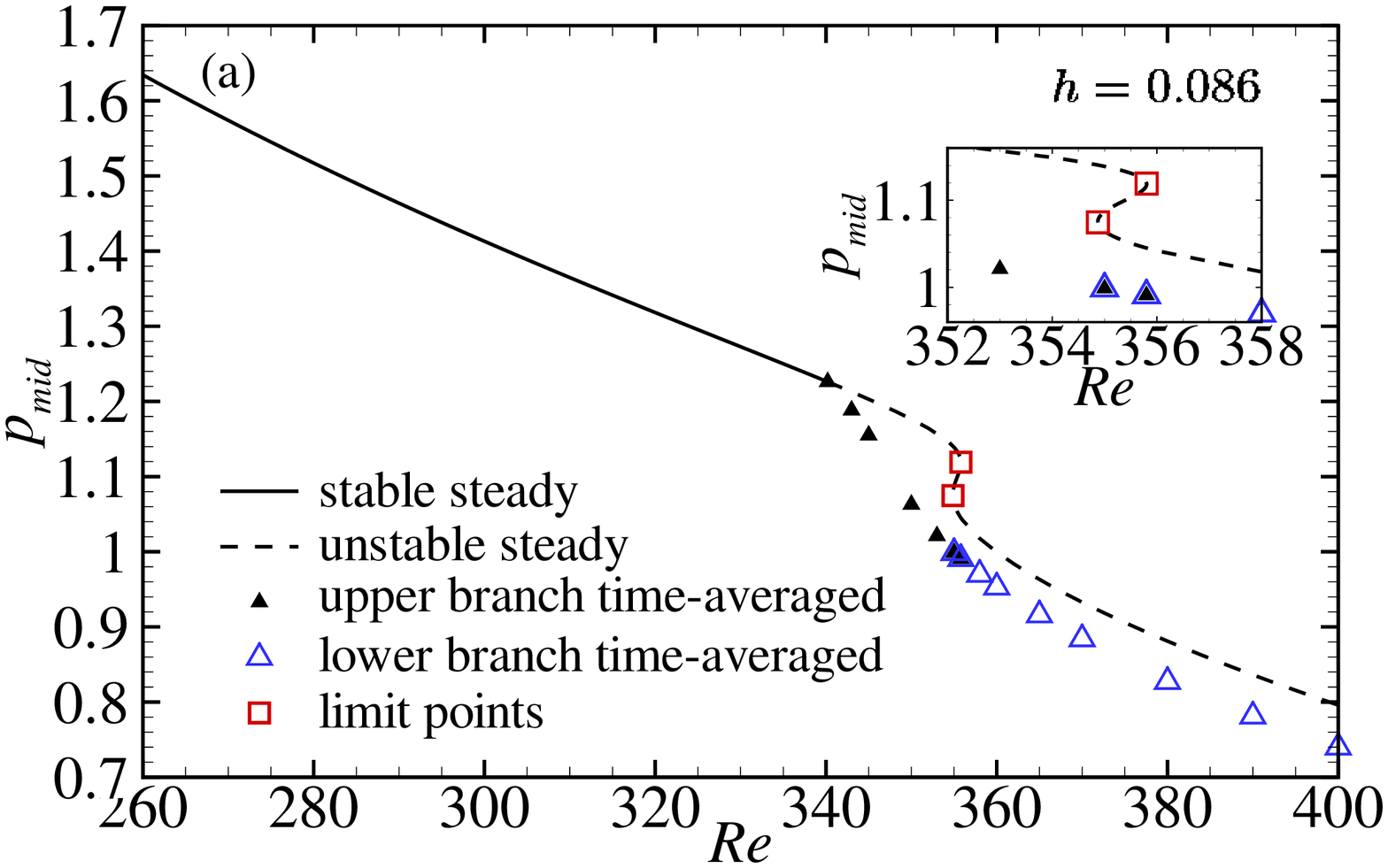}
\includegraphics[width=0.48\textwidth]{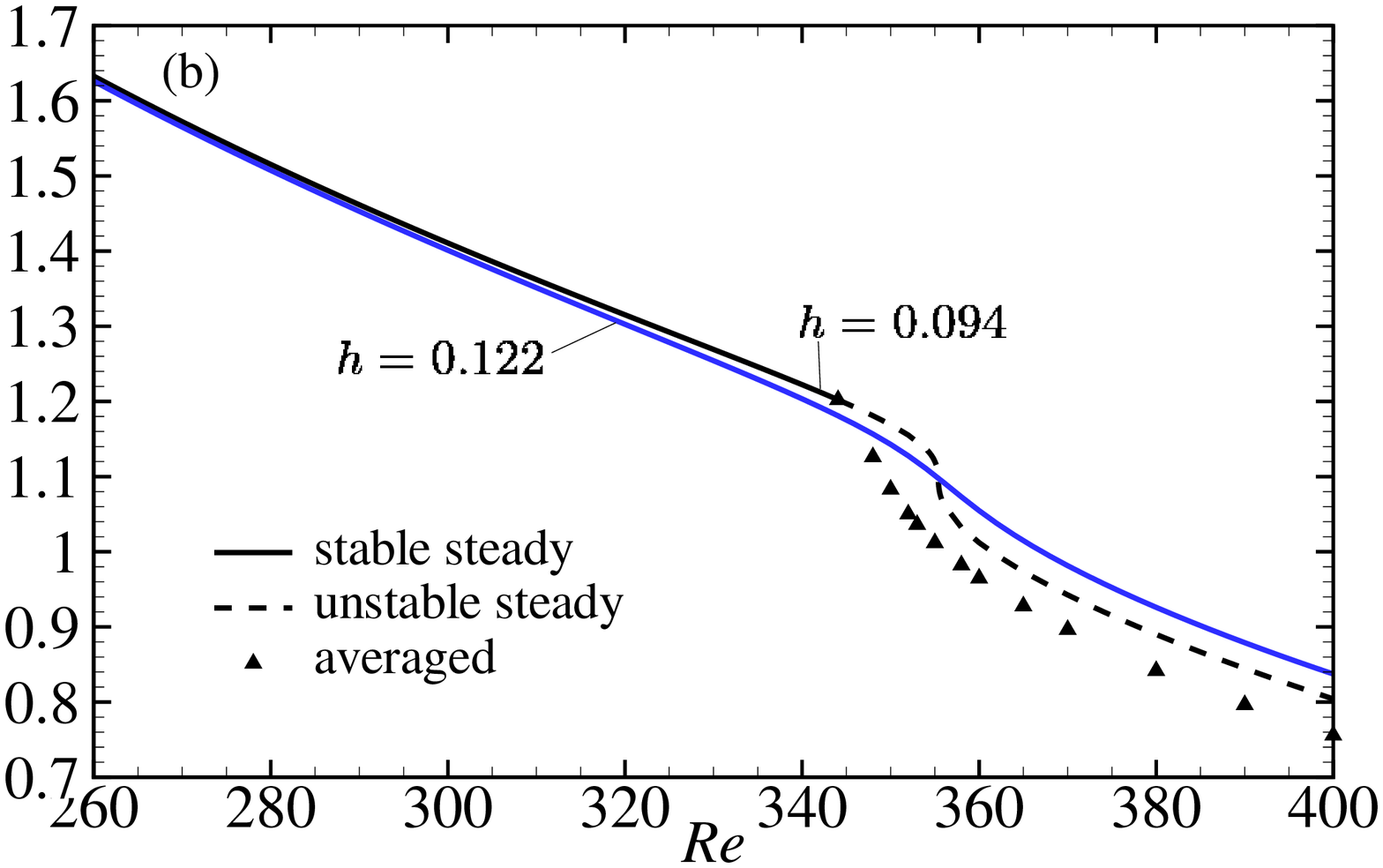}
\caption{Bifurcation diagrams of midpoint wall pressure $p_{mid}$ as a function of the Reynolds number $Re$ at  $T=0$, $p_e=1.1$ for (a) $h=0.086$ and (b) $h=0.094$ and $h=0.122$. The steady midpoint pressure $p_{mid}$ is denoted as solid (stable) and dashed (unstable) lines. The time-averaged midpoint wall pressure $p_{mid}^{avg}$ is denoted as filled triangles, the upper and lower branch limit points are denoted as squares.}\label{Fig:pmid-st-avg-T0-ck1-2}
\end{figure}
To evaluate the role of the beam thickness, $h$, in the unsteady response of the system, Fig. \ref{Fig:pmid-st-avg-T0-ck1-2} presents bifurcation diagrams showing the onset of self-excited oscillations, plotting the time-averaged midpoint pressure alongside the corresponding steady midpoint pressure as a function of the Reynolds number for three beam thicknesses ($h=0.086$, 0.094 and 0.012) with zero pre-tension and fixed external pressure ($p_e=1.1$). 
For beam thickness $h=0.086$ (Fig. \ref{Fig:pmid-st-avg-T0-ck1-2}a), the system exhibits multiple steady states in the range $354.87\lesssim Re\lesssim 355.8$. In this case the oscillations from the upper and lower steady branches merge into one family of oscillations across the region with multiple steady states (similar to Fig. \ref{Fig:pmid-st-avg-Tt0-ck0016}). When the beam thickness is increased to $h=0.094$, the multiple steady states vanish but the system still exhibits transition to self-excited oscillations at $Re\approx344.07$ (denoted as the dashed line), which grow in amplitude as the Reynolds number increases (Fig. \ref{Fig:pmid-st-avg-T0-ck1-2}b). However, as the beam thickness is further increased to $h=0.122$ the system no longer becomes unstable to oscillations and the steady system has a unique solution for all external pressures (Fig. \ref{Fig:pmid-st-avg-T0-ck1-2}b). In summary, increasing the thickness of the beam (and hence the bending stiffness) suppresses the onset of self-excited oscillations as well as the region with multiple steady states.

\subsection{Oscillations with Large Pre-tension}\label{sec:un_largeT}

\begin{figure}[t!]
\centering
\includegraphics[width=0.48\textwidth]{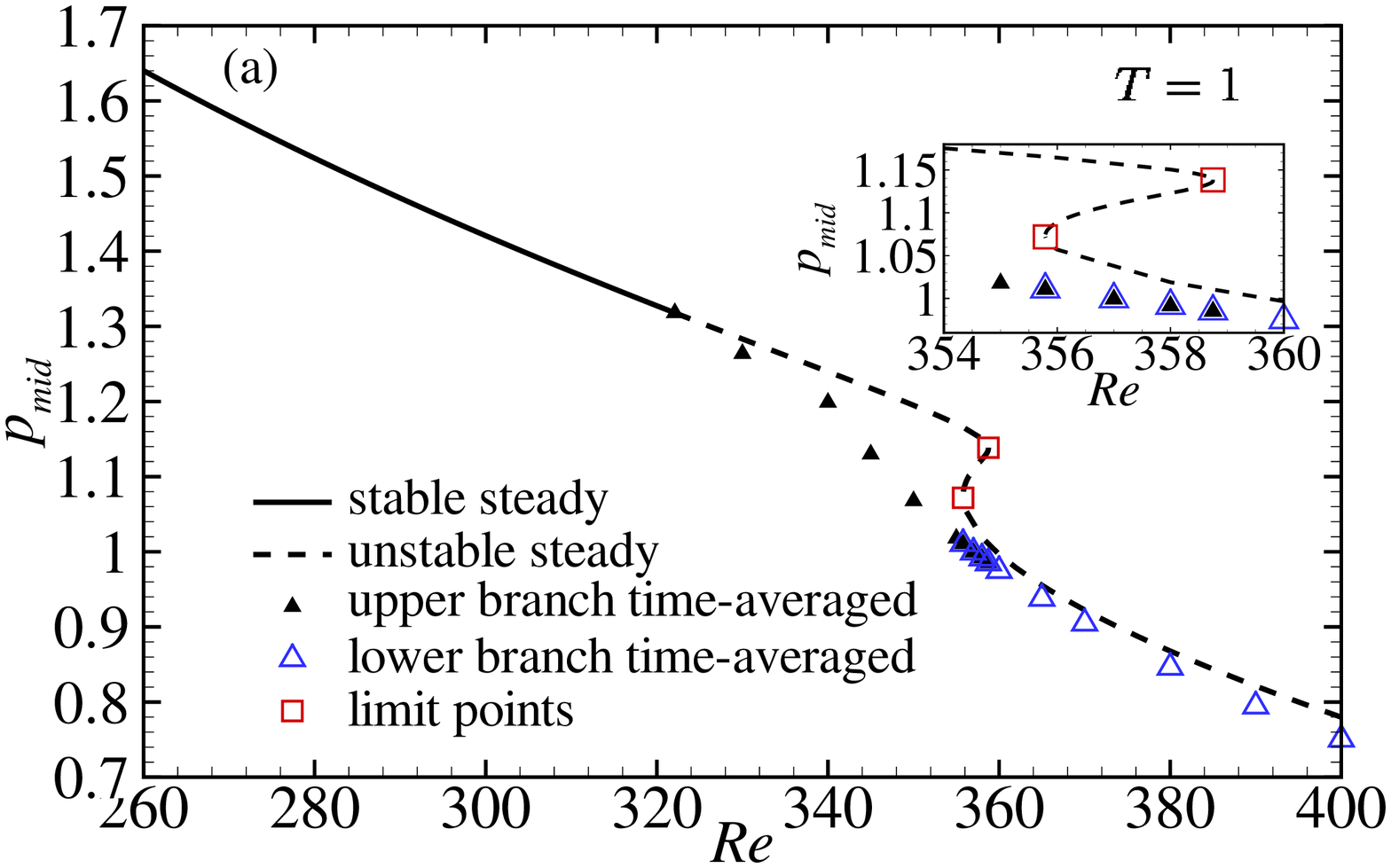}
\includegraphics[width=0.48\textwidth]{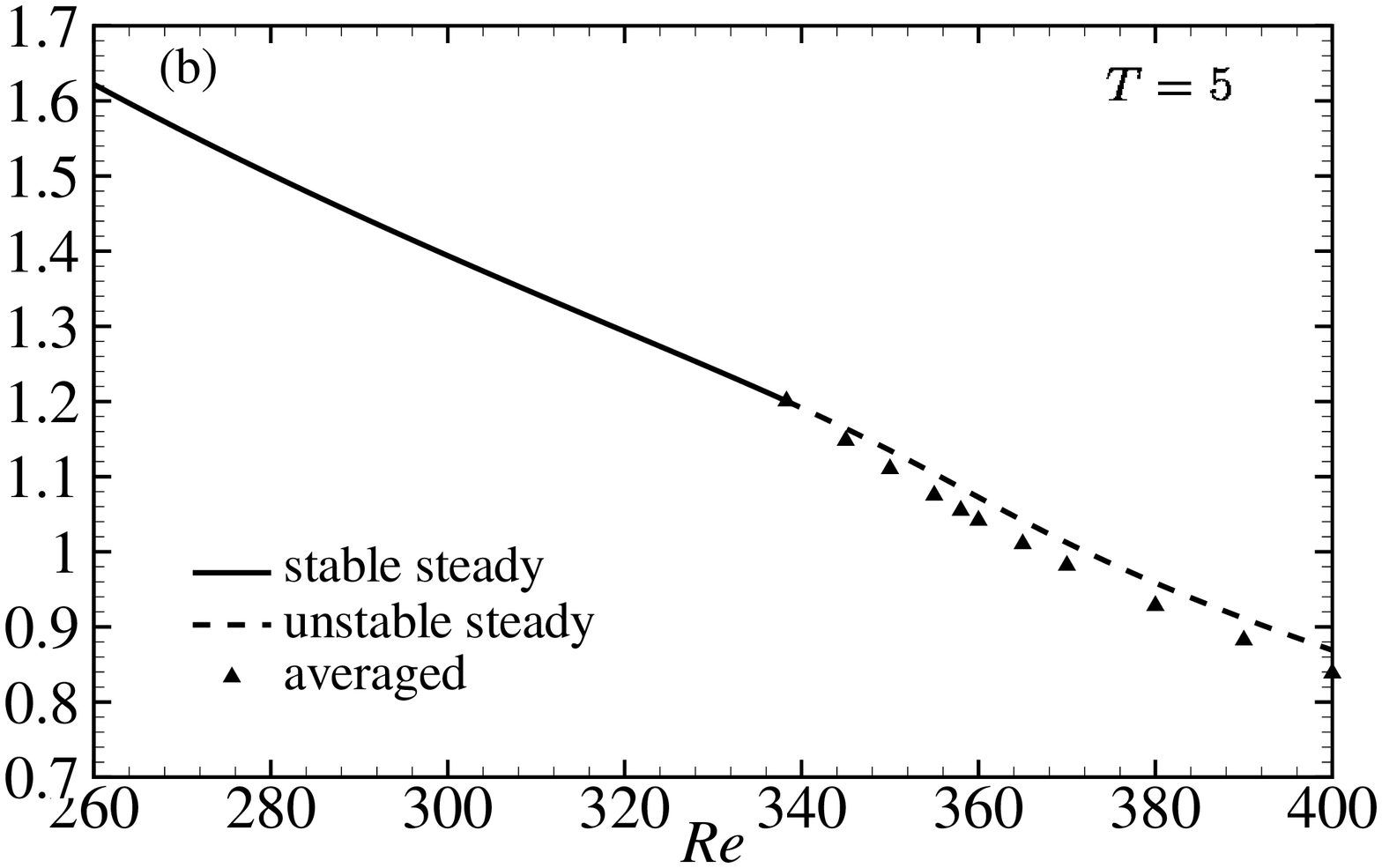}
\caption{Bifurcation diagrams of midpoint wall pressure $p_{mid}$ as a function of the Reynolds number $Re$ at  $h=0.01$, $p_e=1.1$ for (a) $T=1$ and (b) $T=5$. The steady midpoint pressure $p_{mid}$ is denoted as solid (stable) and dashed (unstable) lines. The time-averaged midpoint wall pressure $p_{mid}^{avg}$ is denoted as filled triangles, the upper and lower branch limit points are denoted as squares.}\label{Fig:pmid-st-avg-T1-5}
\end{figure}
In order to evaluate the role of pre-tension $T$ in the onset of self-excited oscillations, Fig. \ref{Fig:pmid-st-avg-T1-5} plots a bifurcation diagram showing the time-averaged midpoint wall pressure compared to the steady midpoint pressure as a function of the Reynolds number for large pre-tension $T=1$ (Fig. \ref{Fig:pmid-st-avg-T1-5}a) and $T=5$ (Fig. \ref{Fig:pmid-st-avg-T1-5}b) for fixed external pressure ($p_e=1.1$) and beam thickness ($h=0.01$).
For $T=1$, the steady system exhibits multiple steady solutions (Fig. \ref{Fig:st_ck0016_fix}) and the oscillations initiated close to the upper and lower steady branches merge into one family of oscillations across the region with multiple steady solutions (Fig. \ref{Fig:pmid-st-avg-T1-5}a).
However, for much larger pre-tension ($T=5$) the system exhibits a unique steady state for all Reynolds numbers but this still becomes unstable to self-excited oscillations at $Re\approx 338.3$. As before, the time-averaged midpoint pressure is lower than the corresponding steady value and decreases with the Reynolds number (Fig. \ref{Fig:pmid-st-avg-T1-5}b). Therefore, the onset of oscillations is preserved with increasing pre-tension despite the loss of multiple steady states.

\section{Discussion}\label{sec:discussion}

In this study we revisit a theoretical model for flow in a planar collapsible channel where the flexible wall is modelled as a pre-stressed elastic beam using a modified nonlinear constitutive law \citep{wang2021energetics}, investigating the influence of the external pressure, beam pre-tension and thickness (a proxy for bending stiffness) on the steady and unsteady behaviour of the system. The model was solved numerically using the finite element method.

Similar to the Starling Resistor experiments \citep{bertram1990mapping,bertram1999flow} and previous models of flow in collapsible channels and tubes \citep{armitstead1996study,heil2004efficient,stewart2017instabilities}, our model predicts that the steady system can exhibit multiple co-existing states, consisting of upper, intermediate and lower steady branches. The model predicts that this region of multiple steady states is suppressed by increasing either the wall pre-stress (Fig.~\ref{Fig:st_ck0016_fix}) or the wall bending stiffness (Fig.~\ref{Fig:st_T0_fix}). In the former the critical point for multiple steady states is postponed to larger Reynolds numbers and lower external pressures as the pre-stress increases (Fig.~\ref{Fig:st_ck0016_fix}d-f), whereas in the latter the critical point does not move significantly but the region with multiple steady states narrows as the bending stiffness increases (Fig.~\ref{Fig:st_T0_fix}b-d). 

Previous studies have indicated that self-excited oscillations can (independently) grow from either the upper and the lower branches of steady solutions \citep{wang2021energetics} in the neighbourhood of the region with multiple steady states (Fig.~\ref{Fig:pe_re_t0-ck0016}, \ref{Fig:pmid-pe148-225}). 
Our model predicts that these two families of oscillations eventually merge into a single family of oscillatory behaviour for sufficiently low external pressure (Fig.~\ref{Fig:pmid-st-avg-Tt0-ck0016}). This new merged family of oscillations retains characteristics of both the upper and lower branch oscillations (Fig.~\ref{Fig:pwall-fullfield}). This new family of oscillations is preserved as the beam pre-stress increases, despite the loss of multiple steady states (Fig.~\ref{Fig:pmid-st-avg-T1-5}). Conversely, this merged family of oscillations is suppressed by increasing the beam thickness (i.e.~increasing the bending stiffness of the beam, Fig. \ref{Fig:pmid-st-avg-T0-ck1-2}).

\section*{Acknowledgements}
\noindent
We gratefully acknowledge funding from the Chinese Scholarship Council (DYW) and UK Engineering and Physical Sciences Research Council grants EP/S020950,   EP/S030875 and EP/N014642 (XYL and PSS).

\def\refname{R\lowercase{eferences}}
\bibliographystyle{ws-ijam}   
\bibliography{ref}         

\end{document}